\documentclass[preprint,aps,showkeys,superscriptaddress]{revtex4}%
\usepackage[latin9]{inputenc}
\usepackage{amsmath}
\usepackage{amssymb}
\usepackage{graphicx}
\usepackage{amsfonts}
\usepackage{subfig}%
\makeatletter
\@ifundefined{textcolor}{}
{
\definecolor{BLACK}{gray}{0}
 \definecolor{WHITE}{gray}{1}
 \definecolor{RED}{rgb}{1,0,0}
 \definecolor{GREEN}{rgb}{0,1,0}
 \definecolor{BLUE}{rgb}{0,0,1}
 \definecolor{CYAN}{cmyk}{1,0,0,0}
 \definecolor{MAGENTA}{cmyk}{0,1,0,0}
 \definecolor{YELLOW}{cmyk}{0,0,1,0}
}
\@ifundefined{textcolor}{}{
\definecolor{BLACK}{gray}{0}
 \definecolor{WHITE}{gray}{1}
 \definecolor{RED}{rgb}{1,0,0}
 \definecolor{GREEN}{rgb}{0,1,0}
 \definecolor{BLUE}{rgb}{0,0,1}
 \definecolor{CYAN}{cmyk}{1,0,0,0}
 \definecolor{MAGENTA}{cmyk}{0,1,0,0}
 \definecolor{YELLOW}{cmyk}{0,0,1,0}
}
\@ifundefined{textcolor}{}{
\definecolor{BLACK}{gray}{0}
 \definecolor{WHITE}{gray}{1}
 \definecolor{RED}{rgb}{1,0,0}
 \definecolor{GREEN}{rgb}{0,1,0}
 \definecolor{BLUE}{rgb}{0,0,1}
 \definecolor{CYAN}{cmyk}{1,0,0,0}
 \definecolor{MAGENTA}{cmyk}{0,1,0,0}
 \definecolor{YELLOW}{cmyk}{0,0,1,0}
}
\@ifundefined{textcolor}{}{
\definecolor{BLACK}{gray}{0}
 \definecolor{WHITE}{gray}{1}
 \definecolor{RED}{rgb}{1,0,0}
 \definecolor{GREEN}{rgb}{0,1,0}
 \definecolor{BLUE}{rgb}{0,0,1}
 \definecolor{CYAN}{cmyk}{1,0,0,0}
 \definecolor{MAGENTA}{cmyk}{0,1,0,0}
 \definecolor{YELLOW}{cmyk}{0,0,1,0}
}
\@ifundefined{textcolor}{}{
\definecolor{BLACK}{gray}{0}
 \definecolor{WHITE}{gray}{1}
 \definecolor{RED}{rgb}{1,0,0}
 \definecolor{GREEN}{rgb}{0,1,0}
 \definecolor{BLUE}{rgb}{0,0,1}
 \definecolor{CYAN}{cmyk}{1,0,0,0}
 \definecolor{MAGENTA}{cmyk}{0,1,0,0}
 \definecolor{YELLOW}{cmyk}{0,0,1,0}
}

\@ifundefined{showcaptionsetup}{}{ \PassOptionsToPackage{caption=false}{subfig}}
\@ifundefined{showcaptionsetup}{}{ \PassOptionsToPackage{caption=false}{subfig}}
\@ifundefined{showcaptionsetup}{}{ \PassOptionsToPackage{caption=false}{subfig}}
\@ifundefined{showcaptionsetup}{}{ \PassOptionsToPackage{caption=false}{subfig}}
\@ifundefined{showcaptionsetup}{}{ \PassOptionsToPackage{caption=false}{subfig}}
\makeatother
\begin{document}
\title{{\Large {}Squeezed coherent states for a free particle with time-varying mass}}
\author{A. S. Pereira}
\email{alfaspereira@gmail.com}
\affiliation{Instituto Federal do Par\'{a}, 68600-000 Bragan\c{c}a, Par\'{a}, Brazil }
\affiliation{Departamento de F\'{\i}sica, Universidade Federal de Campina Grande, Caixa
Postal 10071, 58429-900 Campina Grande, Para\'{\i}ba, Brazil}
\author{A. S. Lemos}
\email{adiellemos@gmail.com}
\affiliation{Departamento de F\'{\i}sica, Universidade Federal de Campina Grande, Caixa
Postal 10071, 58429-900 Campina Grande, Para\'{\i}ba, Brazil}
\author{F. A. Brito}
\email{fabrito@df.ufcg.edu.br}
\affiliation{Departamento de F\'{\i}sica, Universidade Federal de Campina Grande, Caixa
Postal 10071, 58429-900 Campina Grande, Para\'{\i}ba, Brazil}
\affiliation{Departamento de F\'{\i}sica, Universidade Federal da Para\'{\i}ba, Caixa
Postal 5008, 58051-970 Jo\~{a}o Pessoa, Para\'{\i}ba, Brazil}

\pacs{03.65.Sq, 03.65.Fd, 03.65.Ca}

\begin{abstract}
We obtain the squeezed coherent states (SCS) for a free particle with
exponentially time-varying mass. We write these states in terms of the squeeze
and displacement parameters on the time-independent Fock states. Thus, we find
a condition on the displacement parameter such that the SCS can be considered
semiclassical states. We show that it is possible to obtain the coherent
states (CS) for a free particle with minimal uncertainty as long as the mass
increases with the time. We analyze the transition probability of a system
initially prepared in the time-independent Fock states to the free particle SCS.

\end{abstract}
\keywords{Coherent squeezed states, Nonunitary approach, Integrals of motion}\maketitle

\section{Introduction}

It is well-known that the motion of a free particle occurs when the sum of all
external forces is zero, which leads to the simplest physical system that we
know in nature. In this case, it is advantageous to use this model to
introduce some basic concepts, e.g., the wave packets representing a
semiclassical motion and the uncertainty relations \citep{LIT1986,AND2008}, in
introductory quantum mechanics text books. Furthermore, this system can be
seen as a limiting case of the harmonic oscillator when the frequency goes to
zero, which in quantum mechanics produces the correlated coherent states
\citep{SHA1996} with applications in optics \citep{SAL2007}.

The coherent states (CS) are known as quantum states closer to a semiclassical
description for a given system under consideration. These states were obtained
in $1926$ by Schr\"{o}dinger for the harmonic oscillator system
\citep{SCH1926}, which can be found in three equivalent ways: $(i)$ as
eigenstates of the annihilation operator, $(ii)$ as states generated by acting
the displacement operator on a preferred state (vacuum state), and $(iii)$ as
states that minimize the Heisenberg uncertainty relation
\citep{GLA1963,KLA1963,SUD1963}. In turn, squeezed states (SS) form a class of
nonclassical states, which minimize the uncertainty relation. At the same
time, these states present the possibility of reducing the standard deviation
of a physical quantity to a value less than that calculated in the vacuum
state at the expense of increasing the standard deviation in another physical
quantity \citep{BAC2004,SCH2017,DEY2021,ZEL2018}. Both classes of states, CS
and SS, correspond to a particular case of the squeezed coherent states (SCS),
see, e.g., \citep{DOD2003} and the references therein.

Unlike the harmonic oscillator, the free particle has a continuous energy
spectrum, which makes it difficult to introduce the ideas of annihilation and
creation operators. In addition, determining a preferred state for the
application of the displacement operator brings difficulties regarding the
normalization of these states, see \citep{GEL2012}. To circumvent these
problems, some authors address this simple quantum system through the
integrals of motion method, also known as the invariant operator method
\citep{AND1999, BAG2014}. This method consists in determining an operator that
commutes with the Schr\"{o}dinger operator, aiming to obtain the eigenstates
of this integral of motion that must satisfy the Schr\"{o}dinger equation. In
Ref. \citep{GUE2011}, classes of CS and SS have been constructed in terms of
Hermite polynomials using the quantum Arnold transformation. Recently a
detailed study of generalized CS was considered in Ref. \citep{MAA2016},
whereas symplectic tomography were addressed in Ref. \citep{OLG2022}.

Following the recent paper \citep{PER2021} -- in which the SCS are constructed
in a nonunitary approach -- we will obtain the SCS for a free particle with a
mass that varies exponentially with the time. The advantages of the nonunitary
approach compared to the unitary one for the construction of the SCS, lies in
some points: I - identification of the parameters $\xi$ and $\zeta$ (squeeze
and displacement parameters, respectively) with the functions $f$, $g$ and
$\varphi$, whose temporal evolution is determined by the construction of the
integrals of motion; II - as $\hat{S}=\exp\left(  \xi\hat{a}^{\dagger}%
+\frac{1}{2}\zeta\hat{a}^{\dagger2}\right)  $ depends only on $\hat
{a}^{\dagger}$, it becomes advantageous to obtain the condition under which
the SCS satisfy the Schr\"{o}dinger equation. Furthermore, it is direct to
found the relationship between the SCS and the time-independent Fock states,
which is not trivial when considering only the integrals of motion, as
discussed in Refs. \cite{BAG2014,BAG2015}. Following this approach, we will be
able to analyze the role of these parameters in the calculation of uncertainty
relations, as well as find conditions for which the SCS can be considered
semiclassical states. In addition, it will be possible to evaluate the
transition probability from a system initially prepared in a Fock state to the SCS.

This work is structured as follows: In Sec. \ref{Section II}, we construct the
SCS and discuss two limiting cases, i.e., $\zeta=0$, which will represent CS,
while $\xi=0$ describes SS. In Sec. \ref{Section III}, we consider the SCS in
the coordinate representation and get the mean values of the coordinate and
momentum, besides their respective uncertainty relations. In the following, in
Sec. \ref{Section IV}, we find the required conditions to be able to consider
the SCS as a class of semiclassical states. In Sec. \ref{Section V}, we
discuss the probability density and the transition probability for the SCS.
The final remarks are presented in Sec. \ref{Section VI}.

\section{Constructing the SCS of a free particle \label{Section II}}

In this section, we will show the way to be followed for the construction of
the SCS of a free particle. We will start by discussing the free particle
Hamiltonian, and then we will construct the integral of motion, i.e.,
operators that commute with Schr\"{o}dinger operator. Finally, applying the
technique of the recent publication \citep{PER2021}, we will construct the
SCS, which are states annihilated by the integral of motion.

\subsection{Free particle Hamiltonian}

Let us consider the Hamiltonian of a free particle defined on real axis,
$x\in\left(  -\infty,\infty\right)  $, with time-dependent mass $m\left(
t\right)  =m_{0}e^{\gamma t}$, as seen below
\begin{equation}
\hat{H}\left(  t\right)  =\frac{\hat{p}^{2}}{2m\left(  t\right)  },\text{
\ }\hat{p}=-i\hbar\partial_{x},\text{ \ }\partial_{x}=\frac{\partial}{\partial
x},\text{ \ }\left[  \hat{x},\hat{p}\right]  =\hat{x}\hat{p}-\hat{p}\hat
{x}=i\hbar, \label{1}%
\end{equation}
where $\gamma$ is a real parameter with frequency dimension, also known as
damping parameter. The Hamiltonian (\ref{1}) can be taken as a particular case
where the frequency of the parametric oscillator of Caldirola-Kanai goes to
zero \citep{CAL1941,KAN1948}.

In turn, the quantum states $\left\vert \Psi\left(  t\right)  \right\rangle $,
which describe the time evolution of the system, should satisfy the
Schr\"{o}dinger equation
\begin{equation}
\hat{\Lambda}\left(  t\right)  \left\vert \Psi\right\rangle =0,\text{ \ }%
\hat{\Lambda}\left(  t\right)  =\hat{H}\left(  t\right)  -i\hbar\partial_{t},
\label{2}%
\end{equation}
where $\hat{\Lambda}\left(  t\right)  $ is the Schr\"{o}dinger operator. To
simplify the notation, we will omit the time dependence of the analyzed quantities.

Equivalently, we can represent $\hat{H}$ in terms of the annihilation $\hat
{a}$ and creation $\hat{a}^{\dagger}$ bosonic operators, as follows
\begin{equation}
\hat{H}=\frac{\hbar^{2}}{l^{2}m}\frac{\hat{a}^{\dagger}\hat{a}+\hat{a}\hat
{a}^{\dagger}-\hat{a}^{2}-\hat{a}^{\dagger2}}{4}, \label{3}%
\end{equation}
where $l$ is a real parameter with length dimension, and
\begin{align}
&  \hat{a}=\frac{1}{\sqrt{2}}\left(  \frac{\hat{x}}{l}+\frac{il}{\hbar}\hat
{p}\right)  ,\text{ \ }\hat{a}^{\dagger}=\frac{1}{\sqrt{2}}\left(  \frac
{\hat{x}}{l}-\frac{il}{\hbar}\hat{p}\right)  ,\nonumber\\
&  \hat{x}=l\frac{\hat{a}+\hat{a}^{\dagger}}{\sqrt{2}},\text{ \ }\hat{p}%
=\frac{\hbar}{l}\frac{\hat{a}-\hat{a}^{\dagger}}{i\sqrt{2}},\text{ \ }\left[
\hat{a},\hat{a}^{\dagger}\right]  =1. \label{4}%
\end{align}
In turn, these operators act on the Fock states $\left\vert n\right\rangle $
in the form,
\begin{align}
&  \hat{a}\left\vert n\right\rangle =\sqrt{n}\left\vert n-1\right\rangle
,\text{ \ }\hat{a}^{\dagger}\left\vert n\right\rangle =\sqrt{n+1}\left\vert
n+1\right\rangle ,\text{ \ }\hat{n}\left\vert n\right\rangle =\hat{a}%
^{\dagger}\hat{a}\left\vert n\right\rangle =n\left\vert n\right\rangle
,\nonumber\\
&  \hat{a}\left\vert 0\right\rangle =0,\text{ \ }\left\vert n\right\rangle
=\frac{\left(  \hat{a}^{\dagger}\right)  ^{n}}{\sqrt{n!}}\left\vert
0\right\rangle ,\text{ \ }n=0,1,2,...,\nonumber\\
&
%TCIMACRO{\dsum \limits_{n=0}^{\infty}}%
%BeginExpansion
{\displaystyle\sum\limits_{n=0}^{\infty}}
%EndExpansion
\left\vert n\right\rangle \left\langle n\right\vert =1,\text{ \ }\left\langle
n^{\prime}|n\right\rangle =\delta_{n^{\prime},n},\text{ \ }\delta_{n^{\prime
},n}=\left\{
\begin{array}
[c]{c}%
1,\text{ \ if }n^{\prime}=n\\
0,\text{ \ if }n^{\prime}\neq n
\end{array}
\right.  . \label{5}%
\end{align}

In the construction of the SCS of the free particle with time-dependent mass,
we will consider its expansion on the Fock states. In this representation, we
can study the transition probability between harmonic-oscillator-type and the
free particle systems when considering the transition probability of a system
prepared in Fock states to the quantum states of the free particle. This
transition can be seen as the limit at which the oscillator frequency goes to zero.

\subsection{Integrals of motion as a linear combination of the operators
$\hat{a}$ and $\hat{a}^{\dagger}$}

The integrals of motion method, also known as invariant operators, consists of
obtaining operators whose total derivative in time is zero. This condition
allows the construction of eigenstates common to integrals of motion and
Schr\"{o}dinger operator. Furthermore, we have that the eigenvalue of the
integrals of motion must be a complex constant in time. This technique was
introduced by Lewis and Riesenfeld to study the time-dependent harmonic
oscillator \citep{LEW1969}. For more details and applications of this
technique, see, e.g., \citep{DOD1975, BAG2015, PER2022}.

Here, we define integrals of motion in the form of a linear combination of the
operators $\hat{a}$ and $\hat{a}^{\dagger}$, as seen below
\begin{align}
&  \hat{A}=f\hat{a}+g\hat{a}^{\dagger}+\varphi,\nonumber\\
&  \left[  \hat{A},\hat{A}^{\dagger}\right]  =\mu,\text{ \ }\left\vert
f\right\vert ^{2}-\left\vert g\right\vert ^{2}=\left\vert f_{0}\right\vert
^{2}-\left\vert g_{0}\right\vert ^{2}=\mu, \label{6}%
\end{align}
where $f=f\left(  t\right)  $, $g=g\left(  t\right)  $ and $\varphi
=\varphi\left(  t\right)  $ are complex time-dependent functions with the
following initial conditions: $f_{0}=f\left(  0\right)  $, $g_{0}=g\left(
0\right)  $ and $\varphi_{0}=\varphi\left(  0\right)  $. The $\mu$-parameter
is a real constant defined in the range $\mu>0$. This parameter will be used
to rewrite the SCS in terms of the squeeze and displacement parameters, which
will be defined later. In this form, there is a correspondence between the SS
and CS of the group $SU\left(  1,1\right)  $ constructed by Peremolov \citep{PER1986}.

For operator $\hat{A}$ to be an integral of motion, it must satisfy the
following condition,
\begin{equation}
\dot{\hat{A}}\equiv\frac{d\hat{A}}{dt}=\frac{i}{\hbar}\left[  \hat{H},\hat
{A}\right]  +\frac{\partial\hat{A}}{\partial t}=\frac{i}{\hbar}\left[
\hat{\Lambda},\hat{A}\right]  =0, \label{7}%
\end{equation}
At this point, it is interesting to highlight two properties arising from the
Eq. (\ref{7}), which are: $\left(  i\right)  $ The total derivative of
$\hat{A}$ is zero, so the eigenvalue associated with this operator is a
complex constant in time; $\left(  ii\right)  $ Since the integral of motion
commutes with the Schr\"{o}dinger operator, both operators have a common set
of eigenstates. Therefore, we can write
\begin{align}
&  \hat{A}\left\vert \Psi_{z,\lambda}\right\rangle =z\left\vert \Psi
_{z,\lambda}\right\rangle ,\label{7a}\\
&  \hat{\Lambda}\left\vert \Psi_{z,\lambda}\right\rangle =\lambda\left\vert
\Psi_{z,\lambda}\right\rangle , \label{7b}%
\end{align}
where $\lambda=\lambda\left(  t\right)  $ is an arbitrary function on time,
and one can see that $\lambda=\lambda\left(  t\right)  =0$ reproduces the
Schr\"{o}dinger equation (\ref{2}).

Replacing the Eqs. (\ref{3}) and (\ref{6}) into (\ref{7}), we obtain
\begin{equation}
\left(  \frac{\hbar f}{2l^{2}m}+\frac{\hbar g}{2l^{2}m}-i\dot{g}\right)
\hat{a}^{\dagger}-\left(  \frac{\hbar f}{2l^{2}m}+\frac{\hbar g}{2l^{2}%
m}+i\dot{f}\right)  \hat{a}-i\dot{\varphi}=0. \label{8}%
\end{equation}
From (\ref{8}), we must take the coefficients of operators $\hat{a}^{\dagger}$
and $\hat{a}$ to be null, which leads to restrictions on the functions $f$,
$g$, and $\varphi$ in the form
\begin{equation}
\dot{g}=-\frac{i\hbar}{2l^{2}m}\left(  f+g\right)  ,\text{ \ }\dot{f}%
=\frac{i\hbar}{2l^{2}m}\left(  f+g\right)  ,\text{ \ }\dot{\varphi}=0.
\label{9}%
\end{equation}
Note that $\varphi\left(  t\right)  =\varphi\left(  0\right)  =\varphi$. The
first two equations in (\ref{9}) yield the relation $f=-g+g_{0}+f_{0}$ that
substituting it into first Eq. (\ref{9}), we determine the functions $g$, and
$f$, as follows
\begin{align}
&  g=g_{0}-\frac{i\hbar\left(  f_{0}+g_{0}\right)  }{2l^{2}m_{0}}%
\frac{1-e^{-\gamma t}}{\gamma},\nonumber\\
&  f=f_{0}+\frac{i\hbar\left(  f_{0}+g_{0}\right)  }{2l^{2}m_{0}}%
\frac{1-e^{-\gamma t}}{\gamma}. \label{10}%
\end{align}
We must highlight that the operator $\hat{A}$ is parameterized in terms of the
integration constants $f_{0}$ and $g_{0}$. Thus, assuming different values for
these constants will produce new operators $\hat{A}\longrightarrow\hat
{A}\left(  f_{0},g_{0},t\right)  $, i.e., new integrals of motion.

At this point, we can introduce new parameters $\xi=\xi\left(  t\right)  $ and
$\zeta=\zeta\left(  t\right)  $, as seen below
\begin{align}
\xi &  =\frac{\varphi}{f}=\frac{2l^{2}m_{0}\gamma\xi_{0}}{2l^{2}m_{0}%
\gamma+i\hbar\left(  1+\zeta_{0}\right)  \left(  1-e^{-\gamma t}\right)
},\text{ \ }\xi_{0}=\xi\left(  0\right)  =\frac{\varphi}{f_{0}},\label{11a}\\
\zeta &  =\frac{g}{f}=\frac{2l^{2}m_{0}\gamma\zeta_{0}-i\hbar\left(
1+\zeta_{0}\right)  \left(  1-e^{-\gamma t}\right)  }{2l^{2}m_{0}\gamma
+i\hbar\left(  1+\zeta_{0}\right)  \left(  1-e^{-\gamma t}\right)  },\text{
\ }\zeta_{0}=\zeta\left(  0\right)  =\frac{g_{0}}{f_{0}}. \label{11b}%
\end{align}
Thus, we can rewrite $f$ and $\mu$, in the form
\begin{equation}
f=f_{0}\exp\left[  \frac{i\hbar}{2l^{2}}\int_{0}^{t}\frac{1+\zeta\left(
\tau\right)  }{m\left(  \tau\right)  }d\tau\right]  ,\text{ \ }\mu=\left\vert
f\right\vert ^{2}\left(  1-\left\vert \zeta\right\vert ^{2}\right)
=\left\vert f_{0}\right\vert ^{2}\left(  1-\left\vert \zeta_{0}\right\vert
^{2}\right)  . \label{12}%
\end{equation}
For the SCS in the nonunitary approach, the $\xi$,$\zeta$-parameters can be
interpreted as the displacement and squeeze parameters, respectively, see \citep{PER2021}.

\subsection{Time-dependent SCS of a free particle}

In this section, we will construct the SCS in a nonunitary approach. In this
description, one can define the operator $\hat{S}=\exp\left(  \xi\hat
{a}^{\dagger}+\frac{1}{2}\zeta\hat{a}^{\dagger2}\right)  $ and apply a
transformation to the integrals of motion $\hat{A}$ that relates it directly
to the bosonic annihilation operator $\hat{a}$, in the form
\begin{equation}
\hat{a}=\frac{1}{f}\hat{S}\hat{A}\hat{S}^{-1}. \label{13}%
\end{equation}
We find the transformation (\ref{13}) through the Baker--Campbell--Hausdorff
relation $e^{A}Be^{-A}=B+\left[  A,B\right]  +\frac{1}{2}\left[  A,\left[
A,B\right]  \right]  +\ldots$. As we will see, the application of nonunitary
operator $\hat{S}$ simplifies the calculation of $\Phi$-function, which
ensures that the SCS satisfies the Schr\"{o}dinger equation. On the other
hand, in the unitary approach, the operator $\hat{S}$ introduces the operators
$\hat{a}$ and $\hat{a}^{\dagger}$, which requires additional effort in
determining the $\Phi$-function due to the commutation relations.

Considering the annihilation condition of the vacuum state $\hat{a}\left\vert
0\right\rangle =0$, we can write the relation (\ref{13}) in the form
\begin{equation}
\hat{A}\left\vert \xi,\zeta\right\rangle =0, \label{14}%
\end{equation}
where the general form of $\left\vert \xi,\zeta\right\rangle $ is given by
\begin{equation}
\left\vert \xi,\zeta\right\rangle =\Phi\exp\left(  -\xi\hat{a}^{\dagger}%
-\frac{1}{2}\zeta\hat{a}^{\dagger2}\right)  \left\vert 0\right\rangle .
\label{15}%
\end{equation}
Here, $\Phi=\Phi\left(  t\right)  $ is a complex function, which is determined
by imposing that the states $\left\vert \xi,\zeta\right\rangle $ satisfy the
Eq. (\ref{2}), given by
\begin{equation}
\Phi=C\exp\left(  \frac{1}{2}\frac{\zeta^{\ast}\xi^{2}}{1-\left\vert
\zeta\right\vert ^{2}}-\frac{i\hbar}{4l^{2}}\int_{0}^{t}\frac{\zeta+1}{m}%
d\tau\right)  , \label{16}%
\end{equation}
where $C$ is a real constant. See Appendix \ref{Appendix-A}.

The imposition of the normalization condition $\left\langle \zeta,\xi
|\xi,\zeta\right\rangle =1$ allows us to determine the constant $C$. See
Appendix \ref{Appendix-B}. Therefore, the normalized states $\left\vert
\xi,\zeta\right\rangle $, which satisfy Eq. (\ref{2}), have the following
form
\begin{align}
\left\vert \xi,\zeta\right\rangle  &  =\left(  1-\left\vert \zeta
_{0}\right\vert ^{2}\right)  ^{1/4}\exp\left(  \frac{1}{2}\frac{\zeta^{\ast
}\xi^{2}-\left\vert \xi\right\vert ^{2}}{1-\left\vert \zeta\right\vert ^{2}%
}-\frac{i\hbar}{4l^{2}}\int_{0}^{t}\frac{\zeta+1}{m}d\tau\right)  \exp\left(
-\xi\hat{a}^{\dagger}-\frac{1}{2}\zeta\hat{a}^{\dagger2}\right)  \left\vert
0\right\rangle \nonumber\\
&  =\left(  1-\left\vert \zeta_{0}\right\vert ^{2}\right)  ^{1/4}\exp\left(
\frac{1}{2}\frac{\zeta^{\ast}\xi^{2}-\left\vert \xi\right\vert ^{2}%
}{1-\left\vert \zeta\right\vert ^{2}}-\frac{i\hbar}{4l^{2}}\int_{0}^{t}%
\frac{\zeta+1}{m}d\tau\right)
%TCIMACRO{\dsum \limits_{n=0}^{\infty}}%
%BeginExpansion
{\displaystyle\sum\limits_{n=0}^{\infty}}
%EndExpansion
\frac{\left(  -1\right)  ^{n}}{\sqrt{n!}}\left(  \frac{\zeta}{2}\right)
^{\frac{n}{2}}H_{n}\left(  \frac{\xi}{\sqrt{2\zeta}}\right)  \left\vert
n\right\rangle . \label{17}%
\end{align}
In what follows, we will refer the time-dependent states (\ref{17}) as to SCS.

Equivalently, the states $\left\vert \xi,\zeta\right\rangle $ can be obtained
by directly considering the Eq. (\ref{14}). For this, consider the
decomposition of the SCS in terms of the Fock states, in the form
\begin{equation}
\left\vert \xi,\zeta\right\rangle =%
%TCIMACRO{\dsum \limits_{n=0}^{\infty}}%
%BeginExpansion
{\displaystyle\sum\limits_{n=0}^{\infty}}
%EndExpansion
c_{n}\left\vert n\right\rangle , \label{18}%
\end{equation}
where $c_{n}=c_{n}\left(  t\right)  $ are complex time-dependent functions.
Substituting the above expansion into Eq. (\ref{14}), we obtain that
\begin{equation}
c_{n+1}=-\frac{\sqrt{n}}{\sqrt{n+1}}\zeta c_{n-1}-\frac{\xi}{\sqrt{n+1}}c_{n}.
\label{19}%
\end{equation}
From the eq. (\ref{19}) we can write $c_{n}$ in terms of $c_{0}$ employing a
recurrence relation.

We must highlight that obtaining a closed form for these coefficients is not
an easy task. In this sense, the nonunitary approach has shown itself a
practical mechanism to circumvent this problem. Another way to obtain the
closed form of $c_{n}$ is to identify Eq. (\ref{19}) with the difference
equation of the Hermite polynomials, see \citep{ZEL2021}.

Taking into account the form of $c_{n}$, Eq. (\ref{18}), becomes
\begin{equation}
\left\vert \xi,\zeta\right\rangle =c_{0}%
%TCIMACRO{\dsum \limits_{n=0}^{\infty}}%
%BeginExpansion
{\displaystyle\sum\limits_{n=0}^{\infty}}
%EndExpansion
\frac{\left(  -1\right)  ^{n}}{\sqrt{n!}}\left(  \frac{\zeta}{2}\right)
^{\frac{n}{2}}H_{n}\left(  \frac{\xi}{\sqrt{2\zeta}}\right)  \left\vert
n\right\rangle . \label{21}%
\end{equation}
It is interesting to note that $c_{0}$ is equivalent to the function $\Phi$
(see Eq. (\ref{15})), which was introduced as an\textsl{ ad hoc} function,
while $c_{0}$ arises naturally through the recurrence relation.

Applying the normalization condition $\left\langle \xi,\zeta|\xi
,\zeta\right\rangle =1$, and using the Mehler formula for Hermite polynomials
\cite{ERD1953}, we find the following form for $c_{0}$,
\begin{equation}
c_{0}=\left(  1-\left\vert \zeta\right\vert ^{2}\right)  ^{1/4}\exp\left(
-\frac{1}{2}\frac{\left\vert \xi\right\vert ^{2}-\zeta^{\ast}\xi^{2}%
}{1-\left\vert \zeta\right\vert ^{2}}+i\phi\right)  , \label{22}%
\end{equation}
where $\phi=\phi\left(  t\right)  $ must be a real function in order to
guarantee the normalization condition of the states $\left\vert \xi
,\zeta\right\rangle $. Imposing that the states $\left\vert \xi,\zeta
\right\rangle $ satisfy the Schr\"{o}dinger equation (\ref{2}), this implies
that $\phi$ must have the form
\begin{equation}
\phi=-\frac{\hbar}{4l^{2}}\int_{0}^{t}\frac{\operatorname{Re}\left(
\zeta\right)  +1}{m}d\tau. \label{23}%
\end{equation}
Therefore,
\begin{equation}
\left\vert \xi,\zeta\right\rangle =\left(  1-\left\vert \zeta\right\vert
^{2}\right)  ^{1/4}\exp\left(  \frac{1}{2}\frac{\zeta^{\ast}\xi^{2}-\left\vert
\xi\right\vert ^{2}}{1-\left\vert \zeta\right\vert ^{2}}+i\phi\right)
%TCIMACRO{\dsum \limits_{n=0}^{\infty}}%
%BeginExpansion
{\displaystyle\sum\limits_{n=0}^{\infty}}
%EndExpansion
\frac{\left(  -1\right)  ^{n}}{\sqrt{n!}}\left(  \frac{\zeta}{2}\right)
^{\frac{n}{2}}H_{n}\left(  \frac{\xi}{\sqrt{2\zeta}}\right)  \left\vert
n\right\rangle . \label{24}%
\end{equation}
This result is the same as given in (\ref{17}).

The overlap of the states $\left\vert \xi,\zeta\right\rangle $ for different
squeeze and displacement parameters is given by
\begin{align}
&  \left\langle \zeta_{1},\xi_{1}|\xi_{2},\zeta_{2}\right\rangle
=\frac{\left(  1-\left\vert \zeta_{2}\right\vert ^{2}\right)  ^{1/4}\left(
1-\left\vert \zeta_{1}\right\vert ^{2}\right)  ^{1/4}}{\sqrt{1-\zeta_{1}%
^{\ast}\zeta_{2}}}\exp\left(  \frac{1}{2}\frac{2\xi_{1}^{\ast}\xi_{2}%
-\zeta_{2}\xi_{1}^{\ast2}-\zeta_{1}^{\ast}\xi_{2}^{2}}{1-\zeta_{1}^{\ast}%
\zeta_{2}}\right) \nonumber\\
&  \times\exp\left[  \frac{1}{2}\frac{\zeta_{2}^{\ast}\xi_{2}^{2}-\left\vert
\xi_{2}\right\vert ^{2}}{1-\left\vert \zeta_{2}\right\vert ^{2}}+\frac{1}%
{2}\frac{\zeta_{1}\xi_{1}^{\ast2}-\left\vert \xi_{1}\right\vert ^{2}%
}{1-\left\vert \zeta_{1}\right\vert ^{2}}-\frac{i\hbar}{4l^{2}}\int_{0}%
^{t}\frac{\operatorname{Re}\left(  \zeta_{2}-\zeta_{1}\right)  }{m}%
d\tau\right]  . \label{24.1}%
\end{align}
This result shows that the SCS of a free particle are non-orthogonal. However,
these states satisfy to the completeness relation, see Appendix
\ref{Appendix-C},
\begin{align}
&  \int\left\vert \varphi,\zeta\right\rangle \left\langle \zeta,\varphi
\right\vert d^{2}\varphi=1,\nonumber\\
&  d^{2}\varphi=\left(  \pi\mu\right)  ^{-1}du_{1}du_{2},\text{ \ }%
u_{1}=\operatorname{Re}\left(  \varphi\right)  ,\text{ \ }u_{2}%
=\operatorname{Im}\left(  \varphi\right)  . \label{24.2}%
\end{align}
This means that the SCS form an overcomplete basis in Hilbert space
$\mathcal{H}=L^{2}\left(  \mathbb{R}\right)  $ \cite{GAZ2009}. Indeed, the SCS
form a ``continuous'' family of quantum states in the Hilbert space
$\mathcal{H}$, labeled by all points of the complex plane $\mathbb{C}$.

We can get the CS and SS through the SCS by specifying the squeeze and
displacement parameters, as shown below:

a) Taking into account $\zeta=0\Rightarrow g=0$ we find the CS,
\begin{equation}
\left\vert \xi,0\right\rangle =\exp\left(  \frac{i\hbar}{4l^{2}m\gamma}%
-\frac{\left\vert \xi\right\vert ^{2}}{2}\right)  \exp\left(  -\xi\hat
{a}^{\dagger}\right)  \left\vert 0\right\rangle . \label{24a}%
\end{equation}
The Eqs. (\ref{9}) admit non-trivial solutions as long as we have the limit
\begin{equation}
\gamma t\rightarrow\infty, \label{24b}%
\end{equation}
which corresponds to a large mass particle. Thus, the displacement parameter
satisfies the following differential equation,
\begin{equation}
\dot{\xi}=-\frac{i\hbar}{2l^{2}m}\xi\Longrightarrow\xi=\xi_{0}\exp\left(
\frac{i\hbar}{2l^{2}m_{0}}\frac{e^{-\gamma t}-1}{\gamma}\right)  . \label{24c}%
\end{equation}
From (\ref{24.1}), we have that
\begin{equation}
\left\langle 0,\xi_{1}|\xi_{2},0\right\rangle =\exp\left(  \xi_{1}^{\ast}%
\xi_{2}-\frac{1}{2}\left\vert \xi_{2}\right\vert ^{2}-\frac{1}{2}\left\vert
\xi_{1}\right\vert ^{2}\right)  . \label{24c.1}%
\end{equation}
It is easy to see that the CS are normalized by taking $\xi_{1}=\xi_{2}$ in
the above relation.

b) For SS we must impose the condition $\xi=0\Rightarrow\varphi=0$. Thus, we
find the following states,
\begin{equation}
\left\vert 0,\zeta\right\rangle =\left(  1-\left\vert \zeta\right\vert
^{2}\right)  ^{1/4}\exp\left(  i\phi-\frac{1}{2}\zeta\hat{a}^{\dagger
2}\right)  \left\vert 0\right\rangle =\left(  1-\left\vert \zeta\right\vert
^{2}\right)  ^{1/4}\exp\left(  i\phi\right)
%TCIMACRO{\dsum \limits_{n=0}^{\infty}}%
%BeginExpansion
{\displaystyle\sum\limits_{n=0}^{\infty}}
%EndExpansion
\left(  -\zeta\right)  ^{n}\frac{\sqrt{\left(  2n\right)  !}}{2^{n}%
n!}\left\vert 2n\right\rangle . \label{24d}%
\end{equation}
From (\ref{24.1}), we have that
\begin{equation}
\left\langle \zeta_{1},0|0,\zeta_{2}\right\rangle =\frac{\left(  1-\left\vert
\zeta_{2}\right\vert ^{2}\right)  ^{1/4}\left(  1-\left\vert \zeta
_{1}\right\vert ^{2}\right)  ^{1/4}}{\sqrt{1-\zeta_{1}^{\ast}\zeta_{2}}}%
\exp\left[  -\frac{i\hbar}{4l^{2}}\int_{0}^{t}\frac{\operatorname{Re}\left(
\zeta_{2}-\zeta_{1}\right)  }{m}d\tau\right]  . \label{24e}%
\end{equation}
One can easily verify that the SS are normalized by taking $\zeta_{2}%
=\zeta_{1}$. Except for the time evolution, the states (\ref{24d}) coincide
with the CS of the $SU\left(  1,1\right)  $ group when the Bargmann index is
$k=1/4$ and $\zeta\longrightarrow-\zeta^{\prime}$ \citep{PER1986}.

\section{Coordinate representation of the SCS \label{Section III}}

In this section, the SCS will be considered in the coordinate representation,
which will be useful to obtain the required conditions in order to ensure that
these states can be identified as semiclassical states. In addition, the mean
values of the coordinate and momentum will be obtained, as well as their
respective uncertainty relations.

The coordinate representation of the states $\left\vert \xi,\zeta\right\rangle
$ will be found following some steps. First, we obtain the normalized solution
of the vacuum state that we find through the annihilation condition $\left(
\hat{a}\left\vert 0\right\rangle =0\right)  $, and which, in the coordinate
representation, becomes
\begin{align}
&  \hat{a}\Psi_{0}\left(  x\right)  =\frac{l}{\sqrt{2}}\left(  \frac{x}{l^{2}%
}+\frac{\partial}{\partial x}\right)  \Psi_{0}\left(  x\right)  =0,\text{
\ }\Psi_{0}\left(  x\right)  \equiv\left\langle x|0\right\rangle ,\nonumber\\
&  \Psi_{0}\left(  x\right)  =\frac{1}{\sqrt{l\sqrt{\pi}}}\exp\left(
-\frac{x^{2}}{2l^{2}}\right)  ,\text{ \ }\int_{-\infty}^{\infty}\left\vert
\Psi_{0}\left(  x\right)  \right\vert ^{2}dx=1. \label{25}%
\end{align}
Applying $n$ times creation operator $\hat{a}^{\dagger}$ in the state
$\Psi_{0}\left(  x\right)  $, we can determine $\Psi_{n}\left(  x\right)
\equiv\left\langle x|n\right\rangle $, as seen below
\begin{equation}
\Psi_{n}\left(  x\right)  =\frac{\left(  \hat{a}^{\dagger}\right)  ^{n}}%
{\sqrt{n!}}\Psi_{0}\left(  x\right)  =\frac{H_{n}\left(  \frac{x}{l}\right)
}{\sqrt{l\sqrt{\pi}2^{n}n!}}\exp\left(  -\frac{x^{2}}{2l^{2}}\right)  .
\label{26}%
\end{equation}
In what follows, the SCS $\Psi_{\xi,\zeta}\left(  x,t\right)  \equiv
\left\langle x|\xi,\zeta\right\rangle $ in the coordinate representation takes
the form
\begin{equation}
\Psi_{\xi,\zeta}\left(  x,t\right)  =\frac{\left(  1-\left\vert \zeta
\right\vert ^{2}\right)  ^{1/4}}{\sqrt{\sqrt{\pi}l\left(  1-\zeta\right)  }%
}\exp\left[  -\frac{1+\zeta}{2l^{2}\left(  1-\zeta\right)  }\left(
x+\frac{\sqrt{2}l\xi}{1+\zeta}\right)  ^{2}+\frac{\left(  1+\zeta^{\ast
}\right)  \xi^{2}}{2\left(  1+\zeta\right)  \left(  1-\left\vert
\zeta\right\vert ^{2}\right)  }-\frac{\left\vert \xi\right\vert ^{2}}{2\left(
1-\left\vert \zeta\right\vert ^{2}\right)  }+i\phi\right]  . \label{27}%
\end{equation}

Here, we can obtain the CS by applying the condition $\zeta=0\Rightarrow g=0$.
Thus,
\begin{equation}
\Psi_{\xi,0}\left(  x,t\right)  =\frac{1}{\sqrt{\sqrt{\pi}l}}\exp\left[
-\frac{\left(  x+l\sqrt{2}\xi\right)  ^{2}}{2l^{2}}+\frac{\xi^{2}}{2}%
-\frac{\left\vert \xi\right\vert ^{2}}{2}+\frac{i\hbar}{4l^{2}m\gamma}\right]
. \label{28}%
\end{equation}
It is important to highlight that these states satisfy the Schr\"{o}dinger
equation as long as we have the condition $\gamma t\rightarrow\infty$. On the
other hand, the SS are found when we consider $\xi=0\Rightarrow\varphi=0$, so
that
\begin{equation}
\Psi_{0,\zeta}\left(  x,t\right)  =\frac{\left(  1-\left\vert \zeta\right\vert
^{2}\right)  ^{1/4}}{\sqrt{\sqrt{\pi}l\left(  1-\zeta\right)  }}\exp\left(
-\frac{1+\zeta}{1-\zeta}\frac{x^{2}}{2l^{2}}+i\phi\right)  . \label{29}%
\end{equation}

\subsection{Mean values, standard deviations and uncertainty relations}

Now, we will consider the mean values, standard deviation, and uncertainty
relations for the coordinate and momentum. Given the condition (\ref{14}), we
can easily calculate these quantities by rewriting the position and momentum
in terms of the integrals of motion. It follows from (\ref{4}) and (\ref{6})
that
\begin{align}
\hat{a}  &  =\frac{\hat{B}-\zeta\hat{B}^{\dagger}+\zeta\xi^{\ast}-\xi
}{1-\left\vert \zeta\right\vert ^{2}},\text{ \ }\hat{a}^{\dagger}=\frac
{\hat{B}^{\dagger}-\zeta^{\ast}\hat{B}+\zeta^{\ast}\xi-\xi^{\ast}%
}{1-\left\vert \zeta\right\vert ^{2}},\text{ \ }\hat{B}\equiv\frac{1}{f}%
\hat{A},\nonumber\\
\hat{x}  &  =\frac{l}{\sqrt{2}}\frac{\left(  1-\zeta^{\ast}\right)  \hat
{B}+\left(  1-\zeta\right)  \hat{B}^{\dagger}+2\operatorname{Re}\left(
\zeta^{\ast}\xi-\xi\right)  }{1-\left\vert \zeta\right\vert ^{2}},\nonumber\\
\hat{p}  &  =\frac{\hbar}{\sqrt{2}il}\frac{\left(  1+\zeta^{\ast}\right)
\hat{B}-\left(  1+\zeta\right)  \hat{B}^{\dagger}-2i\operatorname{Im}\left(
\zeta^{\ast}\xi+\xi\right)  }{1-\left\vert \zeta\right\vert ^{2}}. \label{30}%
\end{align}
Applying the condition (\ref{14}), we can easily calculate the mean values of
$\hat{x}$ and $\hat{p}$,
\begin{align}
&  \overline{x}=\overline{x}\left(  t\right)  =\left\langle \zeta
,\xi\left\vert \hat{x}\right\vert \xi,\zeta\right\rangle =-\frac{\sqrt
{2}l\operatorname{Re}\left[  \left(  1-\zeta^{\ast}\right)  \xi\right]
}{1-\left\vert \zeta\right\vert ^{2}}=\overline{x}_{0}+\frac{\overline{p}_{0}%
}{m_{0}}\frac{1-e^{-\gamma t}}{\gamma},\nonumber\\
&  \overline{p}=\overline{p}\left(  t\right)  =\left\langle \zeta
,\xi\left\vert \hat{p}\right\vert \xi,\zeta\right\rangle =-\frac{\sqrt{2}%
\hbar}{l}\frac{\operatorname{Im}\left[  \left(  1+\zeta^{\ast}\right)
\xi\right]  }{1-\left\vert \zeta\right\vert ^{2}}=-\frac{\sqrt{2}%
\hbar\operatorname{Im}\left[  \left(  1+\zeta_{0}^{\ast}\right)  \xi
_{0}\right]  }{l\left(  1-\left\vert \zeta_{0}\right\vert ^{2}\right)
}=\overline{p}_{0},\nonumber\\
&  \overline{x}_{0}=-\frac{\sqrt{2}l}{1-\left\vert \zeta_{0}\right\vert ^{2}%
}\operatorname{Re}\left[  \left(  1-\zeta_{0}^{\ast}\right)  \xi_{0}\right]
=-\frac{\sqrt{2}l\left\vert \xi_{0}\right\vert }{1-\left\vert \zeta
_{0}\right\vert ^{2}}\left[  \cos\left(  \theta_{\xi}\right)  -\left\vert
\zeta_{0}\right\vert \cos\left(  \theta_{\xi}-\theta_{\zeta}\right)  \right]
,\text{ \ }\nonumber\\
&  \overline{p}_{0}=-\frac{\sqrt{2}\hbar\left\vert \xi_{0}\right\vert
}{l\left(  1-\left\vert \zeta_{0}\right\vert ^{2}\right)  }\left[  \sin\left(
\theta_{\xi}\right)  +\left\vert \zeta_{0}\right\vert \sin\left(  \theta_{\xi
}-\theta_{\zeta}\right)  \right]  , \label{31}%
\end{align}
where $\zeta_{0}=\left\vert \zeta_{0}\right\vert e^{i\theta_{\zeta}}$ and
$\xi_{0}=\left\vert \xi_{0}\right\vert e^{i\theta_{\xi}}$. In what follows, we
can express the squeeze and displacement parameters in terms of the mean
values (\ref{31}) in the form
\begin{equation}
\xi=-\frac{1+\zeta}{\sqrt{2}}\frac{\overline{x}}{l}-\frac{1-\zeta}{\sqrt{2}%
}\frac{il\overline{p}}{\hbar}. \label{32}%
\end{equation}

Using (\ref{31}), one can easily verify that the mean values $\overline{x}$
and $\overline{p}$ satisfy the Hamilton equations
\begin{equation}
\dot{x}=\frac{\partial H}{\partial p},\text{ \ }\dot{p}=-\frac{\partial
H}{\partial x},\text{ \ }H=\frac{p^{2}}{2m}, \label{32a}%
\end{equation}
where $H$ is the classical Hamiltonian that corresponds to the quantum
Hamiltonian (\ref{1}). Thus, the pair $\overline{x}$ and $\overline{p}$
represents a classical trajectory in the phase-space of the system under
consideration. All such trajectories can be parameterized by the initial
conditions $\overline{x}_{0}$ and $\overline{p}_{0}$.

In its turn, the standard deviations of $\overline{x}$ and $\overline{p}$ are
shown below
\begin{align}
\sigma_{x}  &  =\sigma_{x}\left(  t\right)  =\sqrt{\overline{x^{2}}%
-\overline{x}^{2}}=\frac{l}{\sqrt{2}}\frac{\left\vert 1-\zeta\right\vert
}{\sqrt{1-\left\vert \zeta\right\vert ^{2}}},\nonumber\\
\sigma_{p}  &  =\sigma_{p}\left(  t\right)  =\sqrt{\overline{p^{2}}%
-\overline{p}^{2}}=\frac{\hbar}{\sqrt{2}l}\frac{\left\vert 1+\zeta\right\vert
}{\sqrt{1-\left\vert \zeta\right\vert ^{2}}}. \label{33}%
\end{align}
From here, we can obtain the Heisenberg uncertainty relation
\begin{equation}
\sigma_{x}\sigma_{p}=\frac{\hbar}{2}\frac{\left\vert 1-\zeta\right\vert
\left\vert 1+\zeta\right\vert }{1-\left\vert \zeta\right\vert ^{2}}%
=\frac{\hbar}{2}\sqrt{1+\frac{4\operatorname{Im}^{2}\left(  \zeta\right)
}{\left(  1-\left\vert \zeta\right\vert ^{2}\right)  ^{2}}}. \label{34}%
\end{equation}
The uncertainty relation shows that for any time, the inequality $\sigma
_{x}\sigma_{p}\geq\frac{\hbar}{2}$ is always satisfied. Moreover, the equality
is found when the squeeze parameter is a real quantity, i.e.,
$\operatorname{Im}\left(  \zeta\right)  =0$. Note also that, in the limit
$\zeta\rightarrow\pm1$ we can to obtain values of $\sigma_{x}$ or $\sigma_{p}$
lower than $\sqrt{\hbar/2}$, which characterizes the squeezing property of the SS.

Now, consider the covariance $\sigma_{xp}$,
\begin{equation}
\sigma_{xp}=\frac{1}{2}\left\langle \zeta,\xi\left\vert \left(  \hat
{x}-\overline{x}\right)  \left(  \hat{p}-\overline{p}\right)  +\left(  \hat
{p}-\overline{p}\right)  \left(  \hat{x}-\overline{x}\right)  \right\vert
\xi,\zeta\right\rangle =-\hbar\frac{\operatorname{Im}\left(  \zeta\right)
}{1-\left\vert \zeta\right\vert ^{2}}. \label{34a}%
\end{equation}
In what follows, we can calculate the Robertson-Schr\"{o}dinger uncertainty
relation \citep{ROB1930},
\begin{equation}
\sigma_{x}^{2}\sigma_{p}^{2}-\sigma_{xp}^{2}=\frac{\hbar^{2}}{4}. \label{34b}%
\end{equation}
This result shows that the SCS of a free particle corresponds to a class of
correlated CS \citep{DOD1980}.

In general, the uncertainty relation is not minimized when evaluated for the
free particle. However, at $t=0$, we can obtain a condition on the parameter
$l$ such that the uncertainty relation is minimized. Initially, let us rewrite
the constants $f_{0}$ and $g_{0}$ in the form
\begin{equation}
f_{0}=\cosh\left(  r\right)  ,\text{ \ }g_{0}=\sinh\left(  r\right)
\Rightarrow\zeta_{0}=\tanh\left(  r\right)  ,\text{ \ }-\infty<r<\infty.
\label{35}%
\end{equation}
These conditions imply that $\mu=\left\vert f_{0}\right\vert ^{2}-\left\vert
g_{0}\right\vert ^{2}=1$. Next, taking the standard deviation at $t=0$, one
can rewrite the parameter $l$, as seen below
\begin{equation}
\sigma_{x_{0}}=\sigma_{x}\left(  0\right)  =\frac{l}{\sqrt{2}}\frac
{1-\zeta_{0}}{\sqrt{1-\zeta_{0}^{2}}}\Rightarrow l=\sqrt{2\frac{1+\zeta_{0}%
}{1-\zeta_{0}}}\sigma_{x_{0}}=\sqrt{2}e^{r}\sigma_{x_{0}}, \label{36}%
\end{equation}
where $\sigma_{x_{0}}=\frac{\hbar}{2\sigma_{p_{0}}}$. We will take this
condition of minimization of the uncertainty relation in the next sections.

\begin{figure}[ptb]
\subfloat[]{\includegraphics[width=8cm,height=5cm]{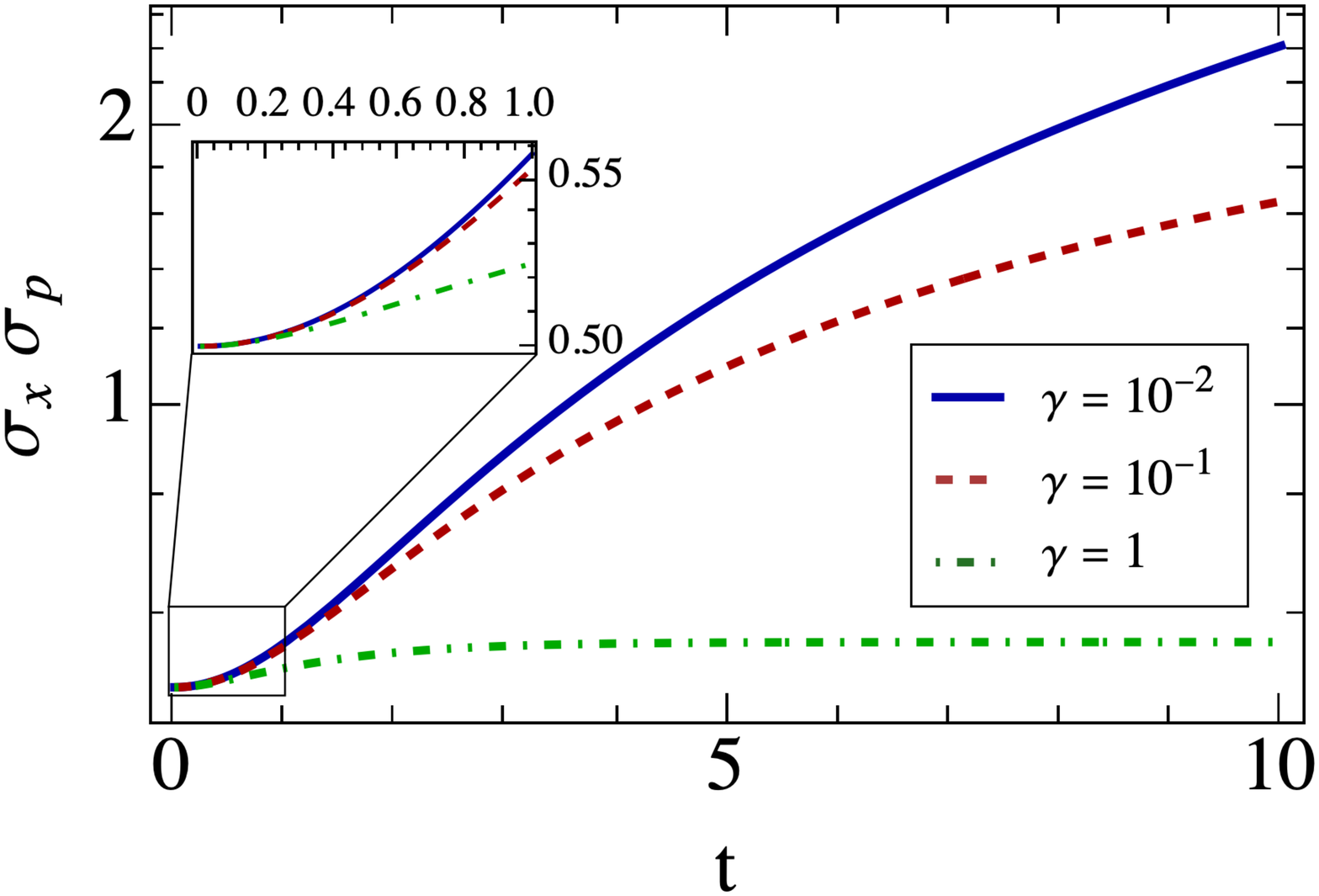} \label{fig1a}
}\hfill{}%
\subfloat[]{\includegraphics[width=8cm,height=5cm]{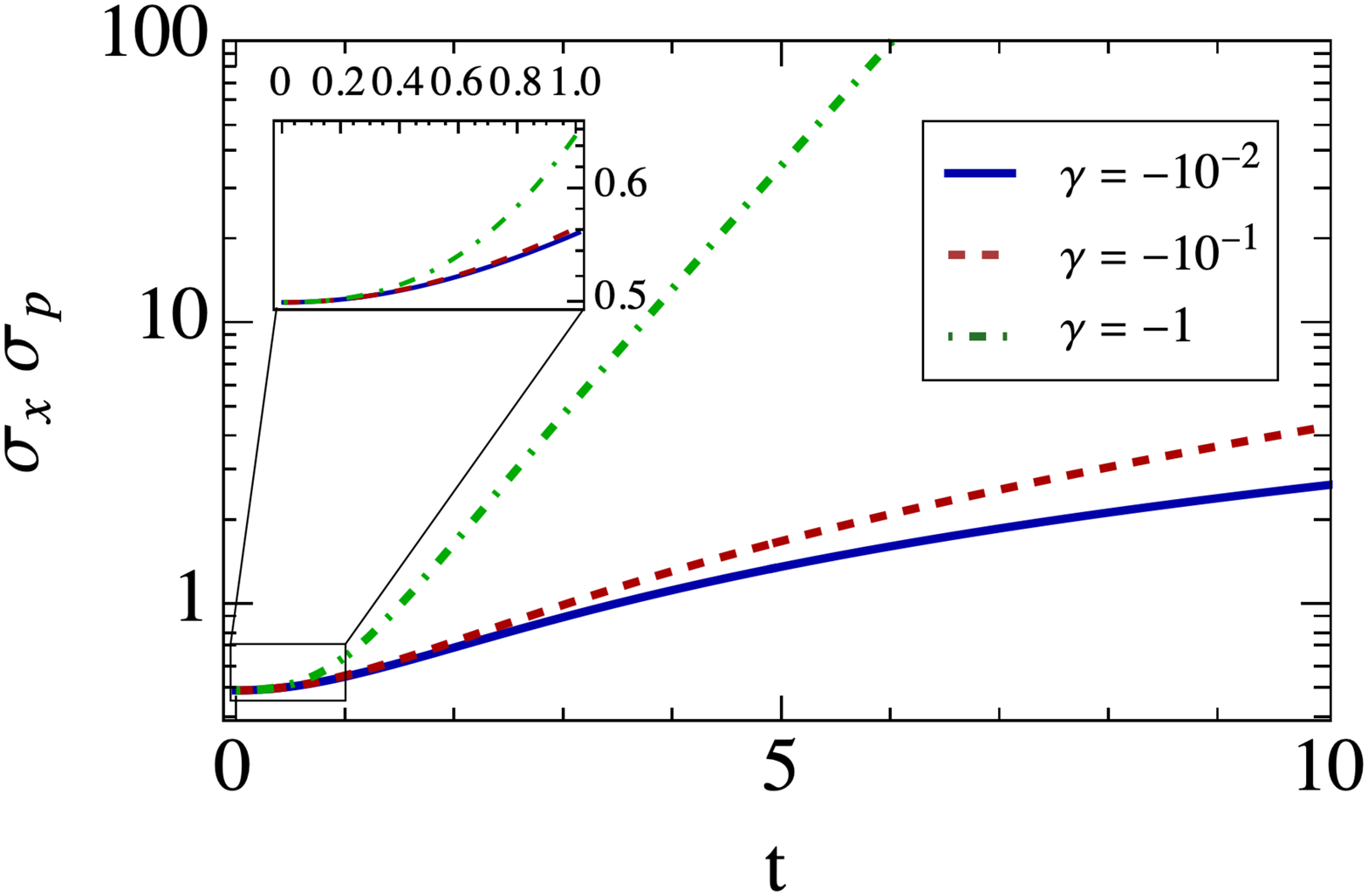} \label{fig1b}
}\caption{The time evolution of free particle Heisenberg uncertainty relation.
We have here fixed $\hbar=1$, $m_{0}=1$, $\sigma_{x_{0}}=1$. We have assumed
$\gamma>0$ in (a), and $\gamma<0$, in (b).}%
\label{fig1}%
\end{figure}

We shall now discuss the time evolution for the Heisenberg uncertainty
relation since we already know $\sigma_{x}$ and $\sigma_{p}$ explicitly.
Calculating the product of the uncertainties on the position and moment, we
get
\begin{equation}
\sigma_{x}\sigma_{p}=\frac{\hbar}{2}\sqrt{1+\frac{\hbar^{2}}{4m_{0}^{2}%
\sigma_{x_{0}}^{4}}\left(  \frac{1-e^{-\gamma t}}{\gamma}\right)  ^{2}}.
\label{eq:55}%
\end{equation}
The Figure \ref{fig1} shows the plot of Eq. (\ref{eq:55}). As we see, if
$\gamma>0$, the Heisenberg uncertainty relation increases with the time and
presents a plateau region (finite value) as $t$ goes to infinity, $\sigma
_{x}\sigma_{p}=\hbar/2\sqrt{1+\hbar^{2}/4m_{0}^{2}\gamma^{2}\sigma_{x_{0}}%
^{4}}$. In the regime $\gamma t\ll1$, we get $\sigma_{x}\sigma_{p}%
=\hbar/2\sqrt{1+\hbar^{2}t^{2}/4m_{0}^{2}\sigma_{x_{0}}^{4}}$, which reproduce
the result obtained in Ref. \citep{LIT1986}. In particular, this equation
shows us the spreading property of the free particle, similarly to what
happens in wave packets. On the other hand, if $\gamma<0$, the corresponding
Heisenberg uncertainty relation goes asymptotically in time as $\sim
e^{\left\vert \gamma\right\vert t}/\gamma$ $\left(  \left\vert \gamma
\right\vert t\rightarrow\infty\right)  $. 

Aiming to discuss the squeeze properties of SCS, we consider the quadratures
$\hat{Q}$ and $\hat{P}$, which one can be obtained by applying the
dimensionless transformation,
\begin{align}
&  \hat{Q}\equiv\hat{x}l^{-1}=\frac{1}{\sqrt{2}e^{r}}\frac{\hat{x}}%
{\sigma_{x_{0}}},\text{ \ }\hat{P}\equiv\frac{l}{\hbar}\hat{p}=\sqrt{2}%
e^{r}\frac{\sigma_{x_{0}}\hat{p}}{\hbar},\text{ \ }\left[  \hat{Q},\hat
{P}\right]  =i,\nonumber\\
&  \tau\equiv\frac{\hbar}{m_{0}l^{2}}t=\frac{\hbar}{2m_{0}e^{2r}\sigma_{x_{0}%
}^{2}}t=e^{-2r}\gamma t\Longrightarrow\gamma t=e^{2r}\tau,\text{ \ }%
\gamma\equiv\frac{\hbar}{2m_{0}\sigma_{x_{0}}^{2}},\nonumber\\
&  \sigma_{Q}=\frac{1}{l}\sigma_{x}=\frac{1}{\sqrt{2}}e^{-r}\sqrt{1+\left(
1-e^{-e^{2r}\tau}\right)  ^{2}},\text{ \ }\sigma_{P}=\frac{l}{\hbar}\sigma
_{p}=\frac{1}{\sqrt{2}}e^{r}.
\end{align}

\begin{figure}[ptb]
\subfloat[]{\includegraphics[width=8cm,height=5cm]{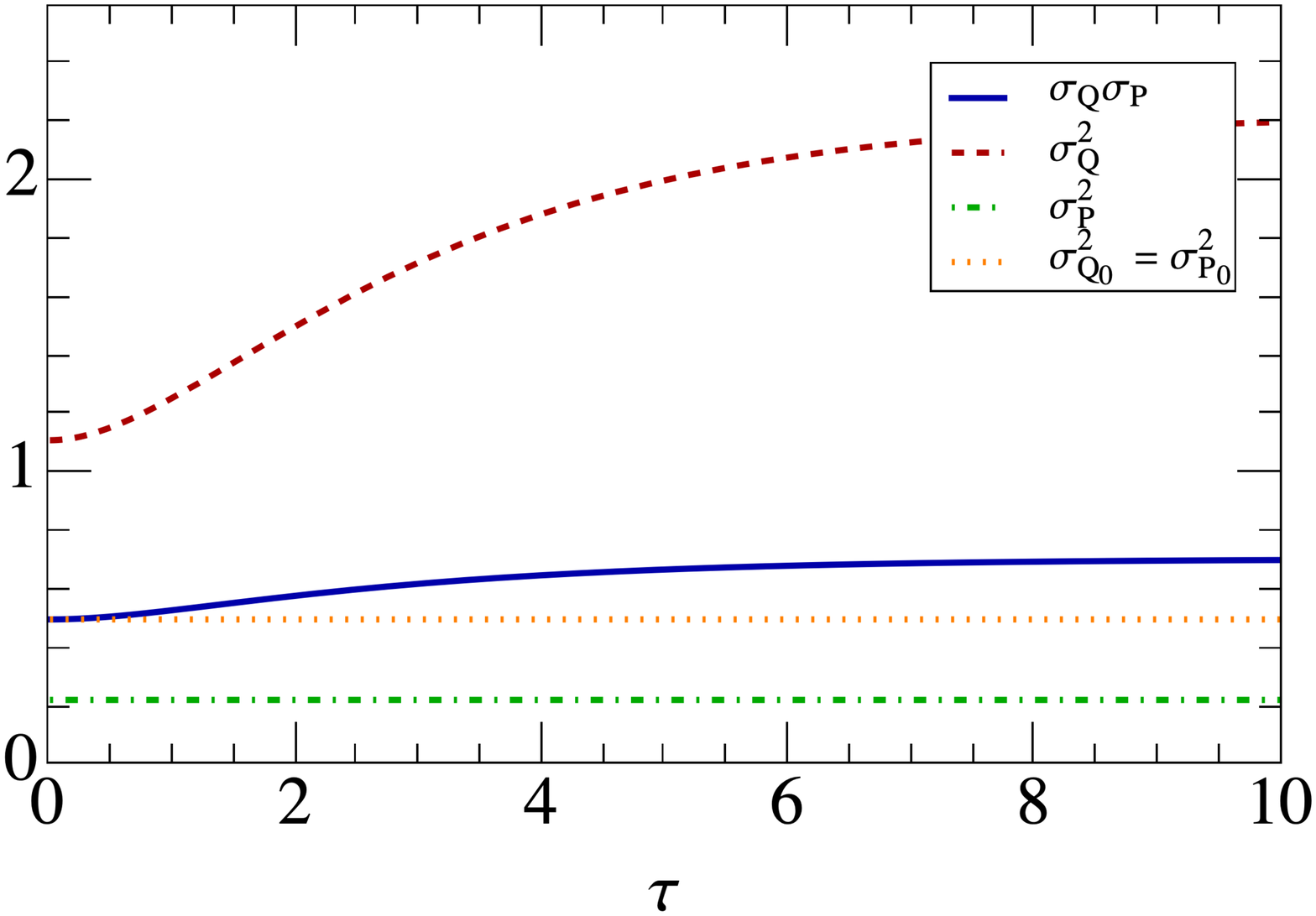} \label{fig2a}
}\hfill{}%
\subfloat[]{\includegraphics[width=8cm,height=5cm]{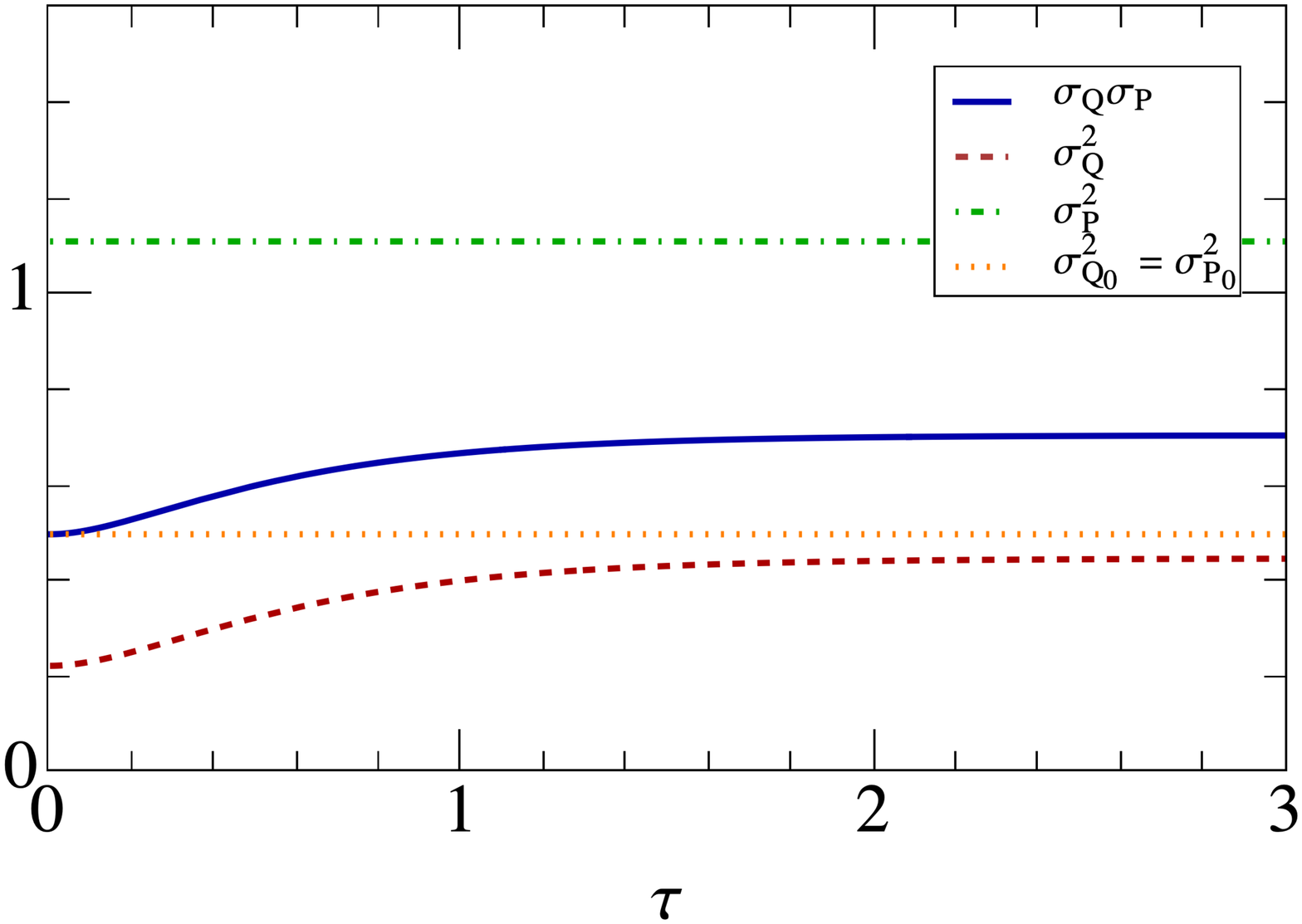} \label{fig2b}
}\caption{Quadrature squeezing was obtained by considering $r=-0.4$ in
\ref{fig2a} and $r=0.4$ in \ref{fig2b}.}%
\label{fig2}%
\end{figure}

An estimate of the nonclassicality of this system is quadrature squeezing,
shown in graphs \ref{fig2}. In Fig. \ref{fig2a}, it is shown that the standard
deviation $\sigma_{P}$ is below the threshold $\sigma_{Q_{0}}^{2}%
=\sigma_{P_{0}}^{2}=1/2$ for values $r<0$ and remains unchanged with time
evolution. On the other hand, for $r>0$, we get that the standard deviation
$\sigma_{Q}$ is below the reference value $1/2$ and goes to a constant value
for $\tau\gg1$, see Fig. \ref{fig2b}. Since the particle has an exponentially
time-dependent mass, $\sigma_{Q}$ grows with time until it reaches a constant
value in all the cases. Besides, in both the cases ($r<0$ and $r>0$), only to
$\tau=0$ we get the minimized uncertainty relation.

\section{SCS as semiclassical states \label{Section IV}}

Here, we will obtain conditions under which we can consider the SCS as a class
of semiclassical states. These states are characterized so that when averaging
physical quantities such as position and momentum, we find that they evolve
along the classical trajectory. In addition, the standard deviation for each
of these means must have a well-behaved range of variation.

In general, the SCS of a free particle cannot be considered semiclassical
states, since the standard deviation in the coordinate increases with time, in
the form
\begin{equation}
\overline{x}=\overline{x}_{0}+\frac{\overline{p}_{0}}{m_{0}}\frac{1-e^{-\gamma
t}}{\gamma},\text{ \ }\sigma_{x}=\sigma_{x_{0}}\sqrt{1+\frac{\hbar^{2}}%
{4m_{0}^{2}\sigma_{x_{0}}^{4}}\left(  \frac{1-e^{-\gamma t}}{\gamma}\right)
^{2}}. \label{37}%
\end{equation}
Semiclassical motion can be characterized by the change slowly with time $t$
in the form of the standard deviation (\ref{37}). Notice that the form of
$\sigma_{x}$ changes owing to the change in the quantity $\frac{\hbar^{2}%
}{4m_{0}^{2}\sigma_{x_{0}}^{4}}\left(  \frac{1-e^{-\gamma t}}{\gamma}\right)
^{2}$ with time. If we assume that the quantity responsible for the change in
$\sigma_{x}$ is significantly smaller than the distance traveled by the
particle in the same amount of time, the form of the standard deviation
(\ref{37}) changes slowly, which characterize, in a certain sense, the
semiclassical motion. This condition is presented as follows:
\begin{equation}
\left\vert \frac{\hbar}{2m_{0}\sigma_{x_{0}}}\frac{1-e^{-\gamma t}}{\gamma
}\right\vert \ll\left\vert \frac{\overline{p}_{0}}{m_{0}}\frac{1-e^{-\gamma
t}}{\gamma}\right\vert , \label{38}%
\end{equation}
or
\begin{equation}
\frac{\hbar}{2\sigma_{x_{0}}}\ll\left\vert \overline{p}_{0}\right\vert .
\label{39}%
\end{equation}
This result coincides with the condition obtained in \citep{BAG2014}. For the
sake of simplicity, let us consider $\overline{p}_{0}$ with $\theta_{\xi}%
=\pi/2$, see Eqs. (\ref{31}), such that we can rewrite the above inequality as
follows
\begin{equation}
\left\vert \varphi\right\vert \gg\frac{1}{2}. \label{40}%
\end{equation}

Given the condition (\ref{40}), we can identify the SCS as semiclassical
states. Substituting the relation (\ref{32}), we can write the semiclassical
states of the free particle in the following form
\begin{align}
&  \Psi_{\overline{x},\overline{p}}^{\sigma_{x}}\left(  x,t\right)
=\frac{\left(  1-\left\vert \zeta\right\vert ^{2}\right)  ^{1/4}}{\sqrt
{\sqrt{2\pi}e^{r}\left(  1-\zeta\right)  \sigma_{x_{0}}}}\exp\left[
-\frac{1+\zeta}{1-\zeta}\frac{\left(  x-\overline{x}\right)  ^{2}}%
{4e^{2r}\sigma_{x_{0}}^{2}}+\frac{i\overline{p}_{0}}{2\hbar}\left(
2x-\overline{x}\right)  +i\phi\right] \nonumber\\
&  =\left(  \frac{1}{2\pi\sigma_{x}^{2}}\frac{1-\zeta^{\ast}}{1-\zeta}\right)
^{1/4}\exp\left[  -\frac{\hbar-2i\sigma_{xp}}{\hbar}\frac{\left(
x-\overline{x}\right)  ^{2}}{4\sigma_{x}^{2}}+\frac{i\overline{p}_{0}}{2\hbar
}\left(  2x-\overline{x}\right)  +i\phi\right]  , \label{41}%
\end{align}
where
\begin{equation}
\zeta=\frac{4m_{0}\gamma\sigma_{x_{0}}^{2}\sinh\left(  r\right)  -i\hbar
e^{-r}\left(  1-e^{-\gamma t}\right)  }{4m_{0}\gamma\sigma_{x_{0}}^{2}%
\cosh\left(  r\right)  +i\hbar e^{-r}\left(  1-e^{-\gamma t}\right)  }.
\label{41a}%
\end{equation}
We must highlight that the condition (\ref{40}) is not valid when considering
the SS since the mean values in the position and momentum are null, which
justifies the SS as a class of nonclassical states. On the other hand, the CS
brings special meaning to this condition, which will be investigated in detail below.

\subsection{CS of a free particle}

Assuming the condition $\zeta=0$, the CS of a free particle takes the form
\begin{equation}
\Psi_{\xi,0}\left(  x,t\right)  =\frac{1}{\sqrt{\sqrt{2\pi}\sigma_{x_{0}}}%
}\exp\left[  -\frac{\left(  x+2\sigma_{x_{0}}\xi\right)  ^{2}}{4\sigma_{x_{0}%
}^{2}}+\frac{\xi^{2}}{2}-\frac{\left\vert \xi\right\vert ^{2}}{2}+\frac
{i\hbar}{8\gamma m\sigma_{x_{0}}^{2}}\right]  . \label{42}%
\end{equation}
Now, knowing that
\begin{align}
&  \xi=-\frac{\overline{x}}{2\sigma_{x_{0}}}-\frac{i\sigma_{x_{0}}\overline
{p}}{\hbar}=\varphi\exp\left(  \frac{i\hbar}{4\sigma_{x_{0}}^{2}m_{0}}%
\frac{e^{-\gamma t}-1}{\gamma}\right)  ,\text{ \ }\varphi=\left\vert
\varphi\right\vert e^{i\theta_{\varphi}},\nonumber\\
&  \overline{x}=-2\sigma_{x_{0}}\operatorname{Re}\left(  \xi\right)
=-2\sigma_{x_{0}}\left\vert \varphi\right\vert \cos\left(  \theta_{\varphi
}+\frac{\hbar}{4m_{0}\sigma_{x_{0}}^{2}}\frac{e^{-\gamma t}-1}{\gamma}\right)
,\nonumber\\
&  \overline{p}=-\frac{\hbar}{\sigma_{x_{0}}}\operatorname{Im}\left(
\xi\right)  =-\frac{\hbar}{\sigma_{x_{0}}}\left\vert \varphi\right\vert
\sin\left(  \theta_{\varphi}+\frac{\hbar}{4m_{0}\sigma_{x_{0}}^{2}}%
\frac{e^{-\gamma t}-1}{\gamma}\right)  , \label{42a}%
\end{align}
we can rewrite the state $\Psi_{\xi,0}\left(  x,t\right)  \longrightarrow
\Psi_{\overline{x},\overline{p}}^{\sigma_{x_{0}}}\left(  x,t\right)  $, as
follows
\begin{equation}
\Psi_{\overline{x},\overline{p}}^{\sigma_{x_{0}}}\left(  x,t\right)  =\frac
{1}{\sqrt{\sqrt{2\pi}\sigma_{x_{0}}}}\exp\left[  -\frac{\left(  x-\overline
{x}\right)  ^{2}}{4\sigma_{x_{0}}^{2}}+\frac{i\overline{p}}{2\hbar}\left(
2x-\overline{x}\right)  +\frac{i\hbar}{8\gamma m\sigma_{x_{0}}^{2}}\right]  .
\label{43}%
\end{equation}
In order for the Eqs. (\ref{9}) to be valid, we must consider the limit
(\ref{24b}), which describes a large mass particle. Also, the displacement
parameter must satisfy the differential Eq. (\ref{24c}). In this case, the
mean values present an oscillatory behavior, indicating that the CS of a free
particle with exponentially time-varying mass mimics the Zitterbewegung effect
\cite{AWO1990}. It is important to note that this effect is not classically
observed {[}see eqs. (\ref{32a}){]}. Furthermore, we can interpret that the
free particle with the time-dependent mass brings out the information from
when was subjected to the action of a restoring force with dispersion. This
situation can be seen as the Caldirola-Kanai type oscillator in the limit that
the frequency goes to zero, which describes a free particle with
time-dependent mass. Furthermore, these states minimize the Heisenberg
uncertainty relation (\ref{34}), as can be seen below
\begin{equation}
\sigma_{x}\sigma_{p}=\frac{\hbar}{2},\text{ \ }\gamma t\rightarrow\infty.
\label{44}%
\end{equation}

\section{Probabilistic analysis \label{Section V}}

In this section, we aim to analyze the behavior of the probability density
$\rho_{\overline{x},\overline{p}}^{\sigma_{x}}\left(  x,t\right)
\equiv\left\vert \Psi_{\overline{x},\overline{p}}^{\sigma_{x}}\left(
x,t\right)  \right\vert ^{2}$ and the transition probability $P_{n}\left(
\zeta,\xi\right)  \equiv\left\vert \left\langle n|\xi,\zeta\right\rangle
\right\vert ^{2}$ for the SCS, and we will present their respective
representations for the semiclassical SCS.

\subsection{Probability density}

We start by considering the time-dependent probability density function given
by
\begin{equation}
\rho_{\overline{x},\overline{p}}^{\sigma_{x}}\left(  x,t\right)  =\frac
{1}{\sqrt{2\pi}\sigma_{x}}\exp\left[  -\frac{\left(  x-\overline{x}\right)
^{2}}{2\sigma_{x}^{2}}\right]  , \label{density}%
\end{equation}
where $\overline{x}$ and $\sigma_{x}$ are given by Eq. (\ref{37}), and
\begin{align}
&  \overline{x}_{0}=-2\sigma_{x_{0}}\left\vert \varphi\right\vert \cos\left(
\theta_{\varphi}\right)  ,\text{ \ }\overline{p}_{0}=-\frac{\hbar\left\vert
\varphi\right\vert }{\sigma_{x_{0}}}\sin\left(  \theta_{\varphi}\right)
,\text{ \ }\sigma_{x_{0}}=\frac{\hbar}{2\sigma_{p_{0}}}. \label{46}%
\end{align}
We see that the Eq. (\ref{density}) describes a Gaussian wave packet centered
in $x=\overline{x}$ with width $\sigma_{x}$. Note that if we assume $\zeta=0$,
we obtain the CS, while $\varphi=0$, we get the SS. Moreover, taking the limit
$\left\vert \varphi\right\vert \gg1/2$, we will find the condition that allows
us to assume the SCS as being semiclassical states.

\begin{figure}[ptb]
\subfloat[]{\includegraphics[width=7cm,height=6cm]{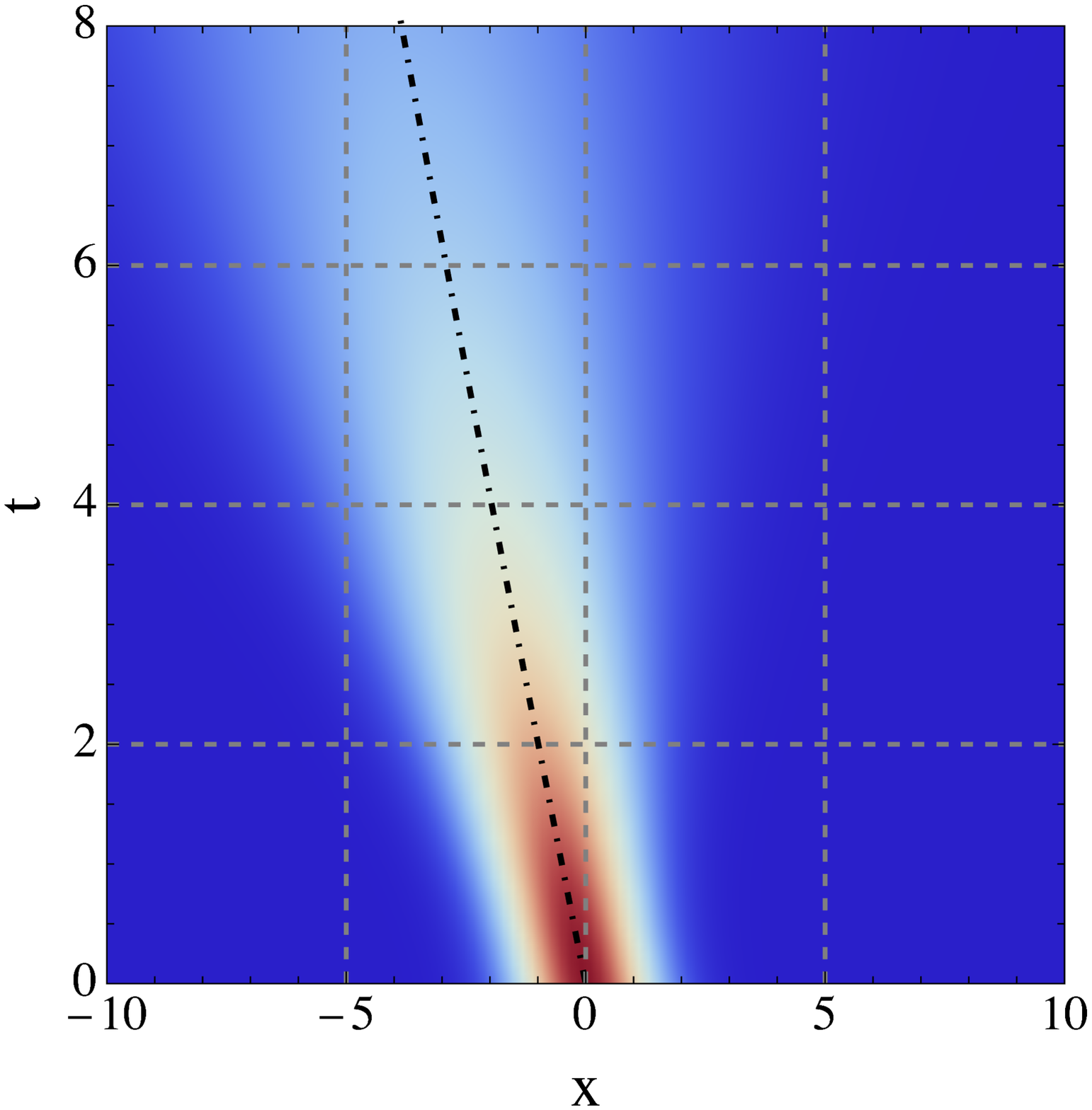} \label{fig3a}
}\hfill{}%
\subfloat[]{\includegraphics[width=8cm,height=6cm]{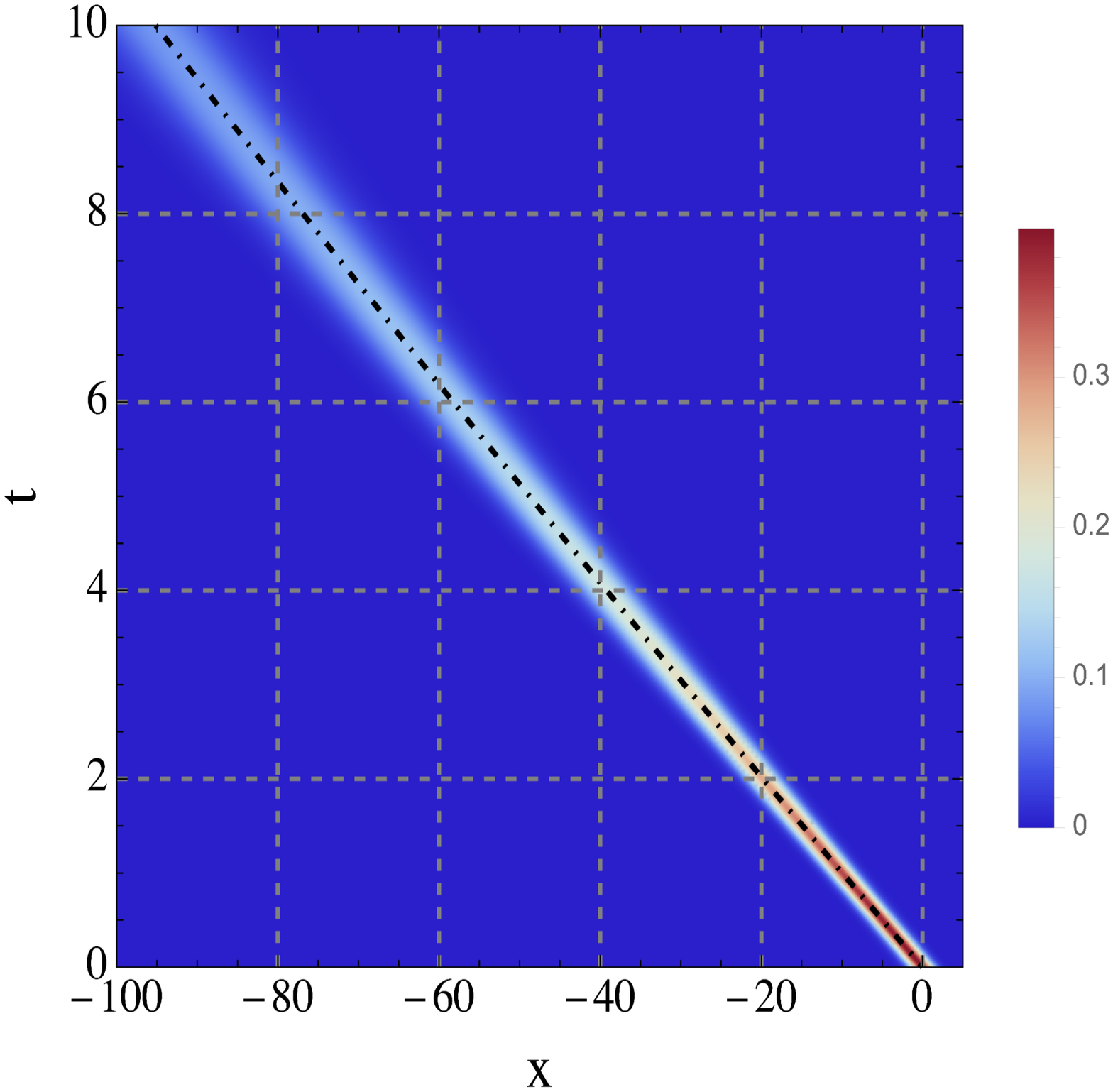} \label{fig3b}
}\caption{The probability density, which evolves in time according to Eq.
(\ref{density}). The dash-dotted curve describes the classical trajectory. We
have assumed $\hbar=1$, $m_{0}=1$, $\sigma_{x_{0}}=1$, and $\gamma=10^{-2}$
with the initial conditions in (a) $\overline{x}_{0}=0$, $\overline{p}%
_{0}=-1/2:\left(  \varphi=1/2\text{ and }\theta_{\varphi}=\pi/2\right)  $ and
in (b) $\overline{x}_{0}=0$, $\overline{p}_{0}=-10:\left(  \varphi=10\text{
and }\theta_{\varphi}=\pi/2\right)  $ . }%
\label{fig3}%
\end{figure}

The time evolution of the probability density has been shown in Fig.
\ref{fig3}. In Fig. \ref{fig3a} at $t=0$ the probability density is localized
around $\overline{x}_{0}=0$ with a initial width $\sigma_{x_{0}}=1$. As the
Figure shows, with time evolution arises the dispersion of the density due to
the spreading in $x$--direction. In Fig. \ref{fig3b} the probability density
plot for a semiclassical SCS $\left(  \varphi\gg1/2\right)  $ has been presented.

\subsection{Transition probability}

Here, we focus on discussing the transition probability. In this case, let us
first admit that the physical system is found in the Fock states and
transitions to another system described by the SCS of a free particle. The
transition probability $P_{n}\left(  \zeta,\xi\right)  \equiv\left\vert
\left\langle n|\xi,\zeta\right\rangle \right\vert ^{2}$ is given as follows
\begin{equation}
P_{n}\left(  \zeta,\xi\right)  =\sqrt{1-\left\vert \zeta\right\vert ^{2}}%
\exp\left[  \frac{\operatorname{Re}\left(  \zeta^{\ast}\xi^{2}\right)
-\left\vert \xi\right\vert ^{2}}{1-\left\vert \zeta\right\vert ^{2}}\right]
\frac{\left\vert \zeta\right\vert ^{n}}{2^{n}n!}\left\vert H_{n}\left(
\frac{\xi}{\sqrt{2\zeta}}\right)  \right\vert ^{2}. \label{transition}%
\end{equation}
In particular with $\zeta=0$, we get
\begin{equation}
P_{n}\left(  0,\xi\right)  =P_{n}\left(  \varphi\right)  =\frac{\left\vert
\varphi\right\vert ^{2n}}{n!}\exp\left(  -\left\vert \varphi\right\vert
^{2}\right)  , \label{probCS}%
\end{equation}
that corresponds to the transition probability for the CS.

\begin{figure}[ptb]
\subfloat[]{\includegraphics[width=8cm,height=5cm]{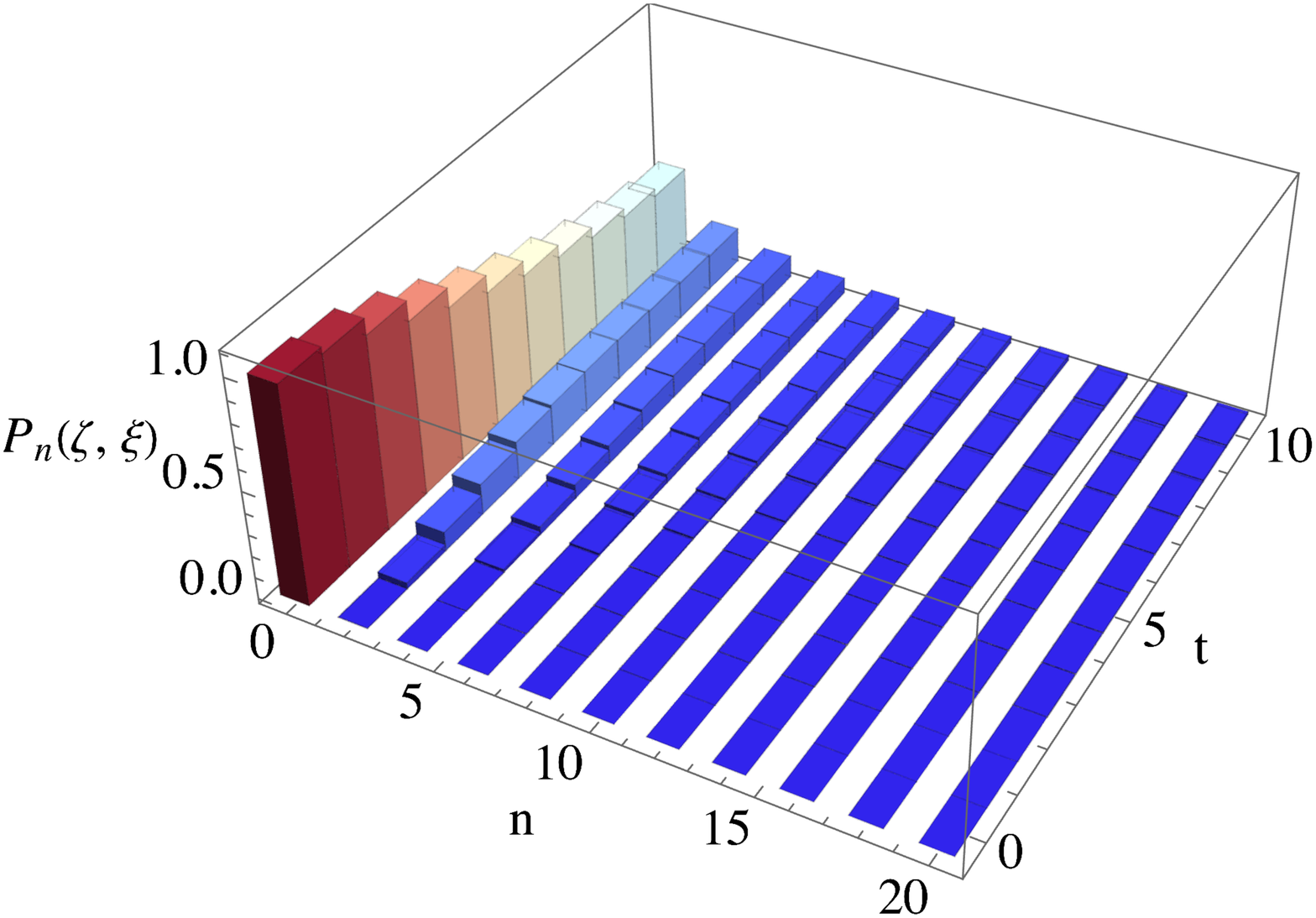} \label{fig4a}
}\hfill{}%
\subfloat[]{\includegraphics[width=8cm,height=5cm]{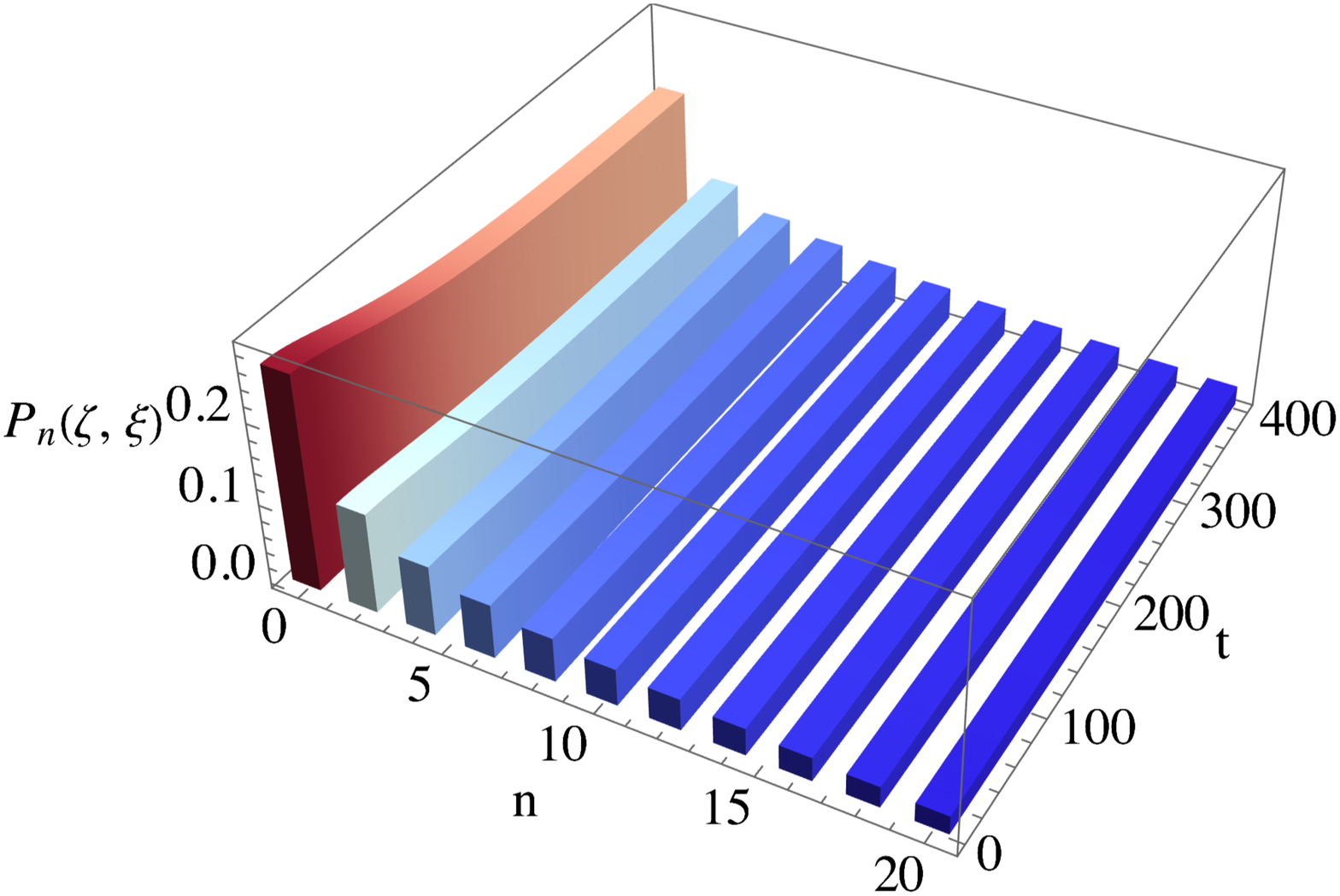} \label{fig4b}
}
\par
\vfill{}
\par
\subfloat[]{\includegraphics[width=8cm,height=5cm]{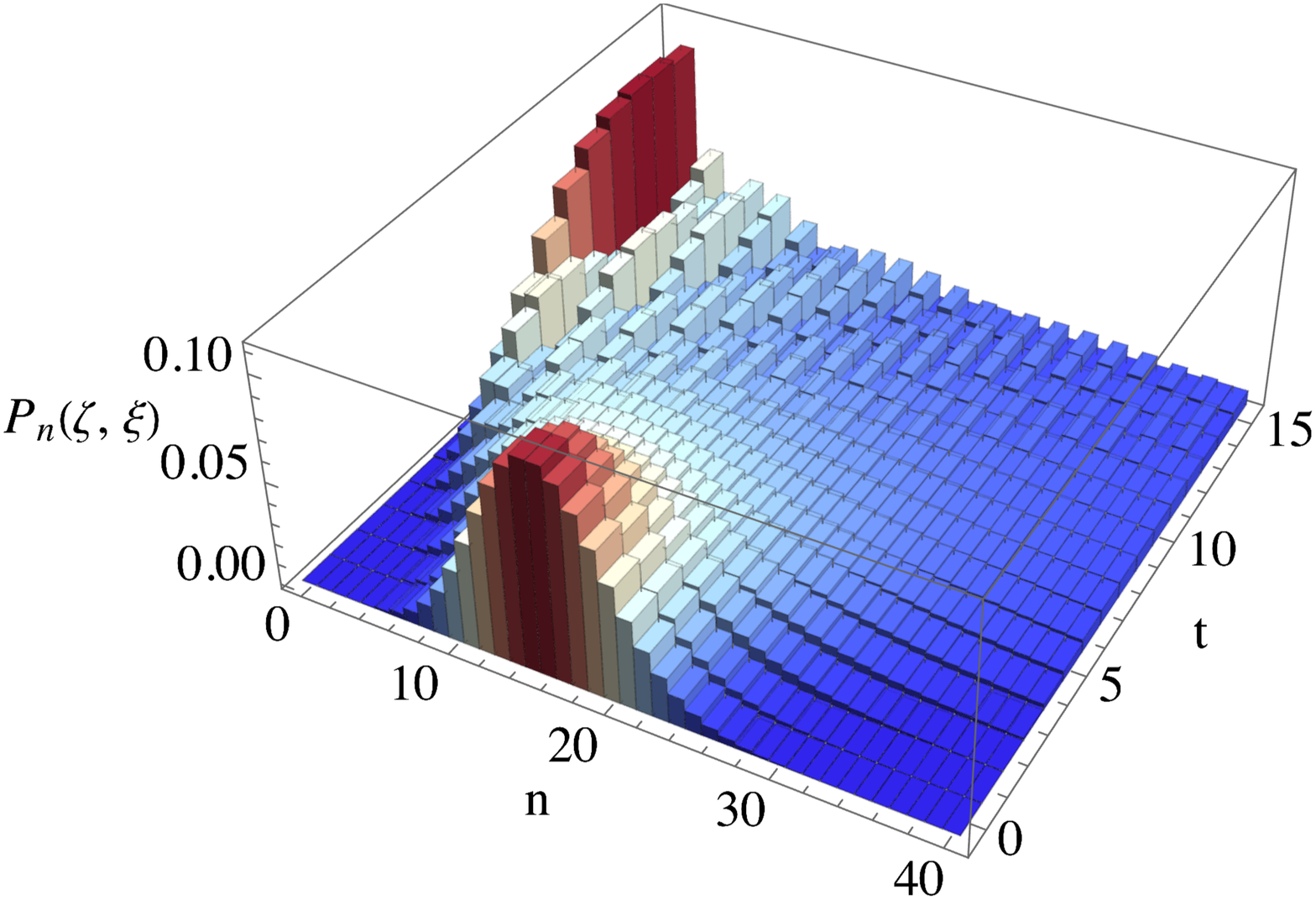} \label{fig4c}
}\hfill{}%
\subfloat[]{\includegraphics[width=8cm,height=5cm]{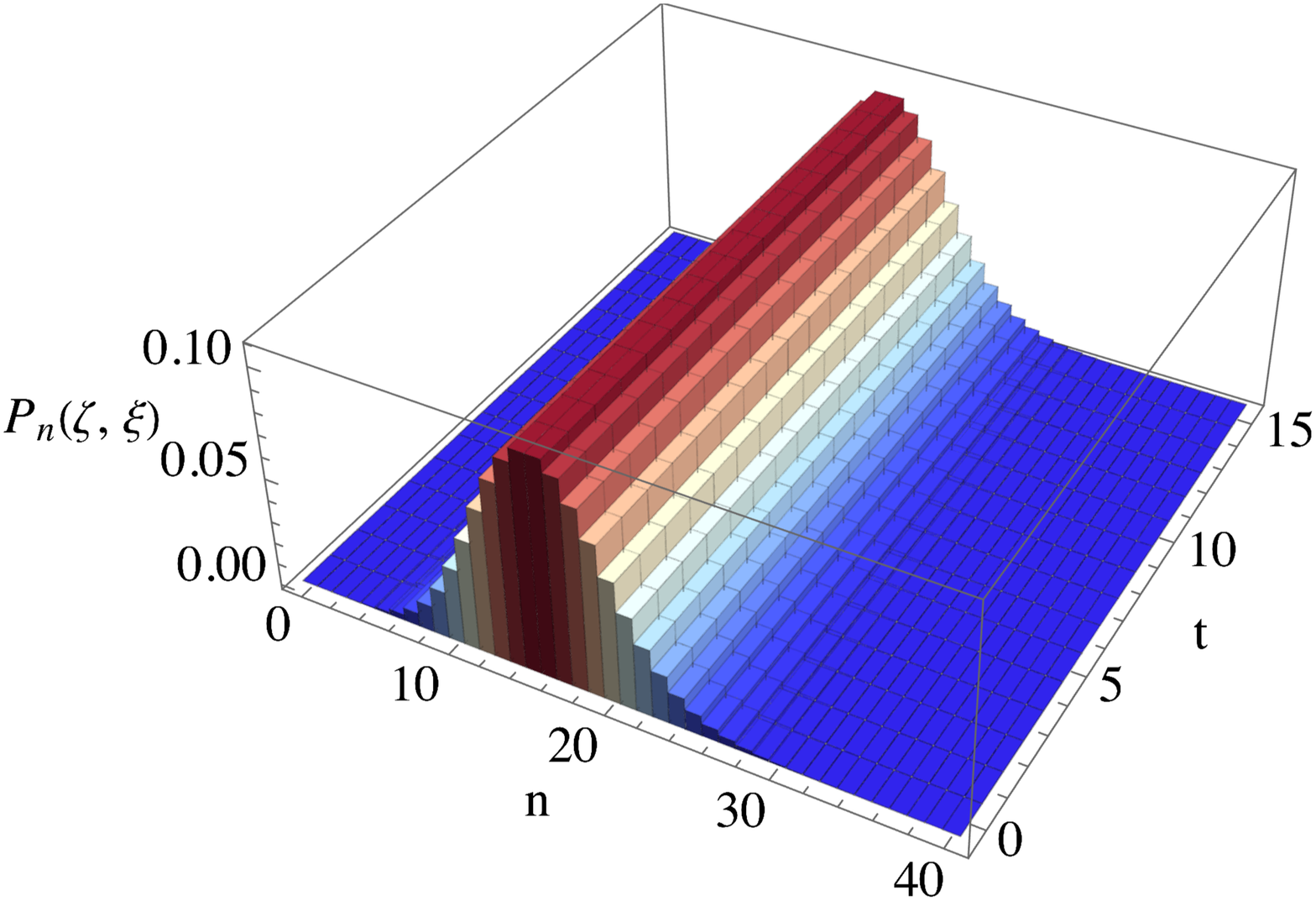} \label{fig4d}
}\caption{The transition probability is shown by considering $\hbar=1$,
$m_{0}=1$, $\sigma_{x_{0}}=1$, $\theta_{\varphi}=\pi/2$, and $\gamma=10^{-2}$.
In (a) and (b) we have kept fixed $\varphi=0$ (SS), and $r=0,2$, respectively.
In (c) has been obtained by assuming $\varphi=4$ (SCS), and $r=0$, while in
(d) we consider $\varphi=4$, and $\zeta=0$ (CS).}%
\label{fig4}%
\end{figure}

As we see by direct comparison between Figs. \ref{fig4a} and \ref{fig4b},
keeping $\varphi$ fixed, the $r$-parameter intensify the transition between
even states. On the other hand, from Fig. \ref{fig4c}, one can see that higher
values for $\varphi$-parameter increase the access to the highest odd states.
Moreover, to higher values of $\varphi\left(  \gg1/2\right)  $, we get the
representation of transition probability for semiclassical SCS, i.e., Fig.
\ref{fig4c}. Notice that the assumed value $r=2$ implies a strong squeeze in
the coordinate. Finally, Fig. \ref{fig4d} corresponds to the transition
probability for the CS $\left(  \zeta=0\right)  $, which is time-independent,
as one can see from Eq. (\ref{probCS}).

\section{Final remarks \label{Section VI}}

In this work, we employ the nonunitary approach to construct the SCS for a
free particle with mass varying exponentially in time. In this approach one
introduces a nonunitary exponential operator composed of a part of the
displacement and squeezed standard operators, in which the squeeze and
displacement parameters are identified with the functions that specify the
integrals of motion. Thus, applying a nonunitary transformation, we related
the integrals of motion directly with the bosonic annihilation operator, which
allowed us to construct the SCS in terms of the time-independent Fock basis.
With this construction, it is straightforward to obtain the condition under
which the SCS satisfies the Schr\"{o}dinger equation.

We reproduce the usual properties such as nonorthogonality, completeness
relation, and probability density of the SCS. Furthermore, we calculated the
mean values, standard deviation, and uncertainty relations. We obtain a
condition for which the Heisenberg uncertainty relation is minimized on the
initial time and determine a condition on the displacement parameter such that
the SCS can be considered semiclassical states. We show that the class of CS
and SS can be obtained from SCS. In particular, to get the CS, which, in turn,
minimizes the Heisenberg uncertainty relation, it is necessary to impose the
limit of a particle with a large mass, which corresponds to the limit $\gamma
t\rightarrow\infty$. We must have this condition in order to guarantee that
$\hat{A}$ are integrals of motion of the Schr\"{o}dinger equation. As
consequence, the mean values present an oscillatory behavior, which one can
associate with the Zitterbewegung effect. As for the SS, we show that they can
be identified with the CS of the $SU\left(  1,1\right)  $ group. Finally, the
transition probability for the SCS, CS, and SS are discussed for several
particular cases.

\begin{acknowledgments}
We would like to thank CNPq, CAPES and CNPq/PRONEX/FAPESQ-PB (Grant No.
165/2018), for partial financial support. ASL and FAB acknowledges support
from CNPq (Grant Nos. 150601/2021-2,312104/2018-9). ASL acknowledges support
from CAPES (Grant No 88887.800922/2023-00). ASP thanks the support of the
Instituto Federal do Par\'{a}.
\end{acknowledgments}

\textbf{Data Availability Statement}

Data sharing not applicable to this article as no datasets were generated or
analyzed during the current study.

\appendix
%dummy comment inserted by tex2lyx to ensure that this paragraph is not empty%dummy comment inserted by tex2lyx to ensure that this paragraph is not empty%dummy comment inserted by tex2lyx to ensure that this paragraph is not empty%dummy comment inserted by tex2lyx to ensure that this paragraph is not empty

\section{Determination of $\Phi$--function \label{Appendix-A}}

We aim to get the $\Phi$--function present in the states (\ref{15}). For this,
we must substitute the states $\left\vert \xi,\zeta\right\rangle $ in the
Schr\"{o}dinger equation, and multiply it by $\left\langle \zeta
,\xi\right\vert $, to have the following differential equation for $\Phi$,
\begin{equation}
\frac{\dot{\Phi}}{\Phi}=\frac{1}{i\hbar}\frac{\left\langle \xi,\zeta\left\vert
\hat{H}\right\vert \xi,\zeta\right\rangle }{\left\langle \xi,\zeta|\xi
,\zeta\right\rangle }+\dot{\xi}\frac{\left\langle \xi,\zeta\left\vert \hat
{a}^{\dagger}\right\vert \xi,\zeta\right\rangle }{\left\langle \xi,\zeta
|\xi,\zeta\right\rangle }+\frac{1}{2}\dot{\zeta}\frac{\left\langle \xi
,\zeta\left\vert \hat{a}^{\dagger2}\right\vert \xi,\zeta\right\rangle
}{\left\langle \xi,\zeta|\xi,\zeta\right\rangle }. \label{A1}%
\end{equation}
The derivative of parameters $\zeta$ and $\xi$ is found by assuming the
definitions (\ref{11a}) and (\ref{11b}) together with Eqs. (\ref{9}), as
follows
\begin{equation}
\dot{\zeta}=\frac{d}{dt}\left(  \frac{g}{f}\right)  =-\frac{i\hbar}{2l^{2}%
m}\left(  1+\zeta\right)  ^{2},\text{ \ }\dot{\xi}=\frac{d}{dt}\left(
\frac{\varphi}{f}\right)  =-\frac{i\hbar}{2l^{2}m}\left(  1+\zeta\right)  \xi.
\label{A2}%
\end{equation}

The mean value of the operators $\hat{H}\left(  t\right)  $, $\hat{a}%
^{\dagger}$ and $\hat{a}^{\dagger2}$ can be easily obtained by considering
these operators in terms of the integrals of motion (\ref{6}), in the form
\begin{align}
&  \hat{a}^{\dagger}=\frac{\hat{B}^{\dagger}-\zeta^{\ast}\hat{B}}{1-\left\vert
\zeta\right\vert ^{2}}+\frac{\zeta^{\ast}\xi-\xi^{\ast}}{1-\left\vert
\zeta\right\vert ^{2}},\text{ \ }\hat{B}\equiv\frac{1}{f}\hat{A},\nonumber\\
&  \hat{a}^{\dagger2}=\frac{\hat{B}^{\dagger2}+\zeta^{\ast2}\hat{B}^{2}%
-2\zeta^{\ast}\hat{B}^{\dagger}\hat{B}+2\left(  \zeta^{\ast}\xi-\xi^{\ast
}\right)  \hat{B}^{\dagger}-2\left(  \zeta^{\ast}\xi-\xi^{\ast}\right)
\zeta^{\ast}\hat{B}}{\left(  1-\left\vert \zeta\right\vert ^{2}\right)  ^{2}%
}+\frac{\left(  \zeta^{\ast}\xi-\xi^{\ast}\right)  ^{2}}{\left(  1-\left\vert
\zeta\right\vert ^{2}\right)  ^{2}}-\frac{\zeta^{\ast}}{1-\left\vert
\zeta\right\vert ^{2}},\nonumber\\
&  \hat{H}\left(  t\right)  =\hbar^{2}\frac{2\left[  1+\left\vert
\zeta\right\vert ^{2}+2\operatorname{Re}\left(  \zeta\right)  \right]  \hat
{B}^{\dagger}\hat{B}-\left(  1+\zeta^{\ast}\right)  ^{2}\hat{B}^{2}-\left(
1+\zeta\right)  ^{2}\hat{B}^{\dagger2}}{4l^{2}m\left(  1-\left\vert
\zeta\right\vert ^{2}\right)  ^{2}}\nonumber\\
&  +\frac{i\hbar^{2}\operatorname{Im}\left(  \zeta\xi^{\ast}-\xi\right)
\left[  \left(  1+\zeta\right)  \hat{B}^{\dagger}-\left(  1+\zeta^{\ast
}\right)  \hat{B}\right]  }{l^{2}m\left(  1-\left\vert \zeta\right\vert
^{2}\right)  ^{2}}+\frac{\hbar^{2}\left\vert \zeta\xi^{\ast}-\xi\right\vert
^{2}}{2l^{2}m\left(  1-\left\vert \zeta\right\vert ^{2}\right)  ^{2}%
}\nonumber\\
&  +\frac{\hbar^{2}\operatorname{Re}\left[  \left(  \zeta+\left\vert
\zeta\right\vert ^{2}\right)  \left(  1-\left\vert \zeta\right\vert
^{2}\right)  -\left(  \zeta\xi^{\ast}-\xi\right)  ^{2}\right]  }%
{2l^{2}m\left(  1-\left\vert \zeta\right\vert ^{2}\right)  ^{2}}+\frac
{\hbar^{2}}{4l^{2}m}.\label{A3}%
\end{align}

Substituting the relations (\ref{A3}) together with the condition (\ref{14}),
we can rewrite Eq. (\ref{A1}) as follows
\begin{equation}
\frac{\dot{\Phi}}{\Phi}=\frac{i\hbar}{4l^{2}m}\left(  \xi^{2}-\zeta-1\right)
=\frac{1}{2}\frac{d}{dt}\left(  \frac{\zeta^{\ast}\xi^{2}}{1-\left\vert
\zeta\right\vert ^{2}}-\frac{i\hbar}{2l^{2}}\int_{0}^{t}\frac{\zeta+1}{m}%
d\tau\right)  . \label{A4}%
\end{equation}
Therefore, the general solution of (\ref{A1}) is given by
\begin{equation}
\Phi=C\exp\left(  \frac{1}{2}\frac{\zeta^{\ast}\xi^{2}}{1-\left\vert
\zeta\right\vert ^{2}}-\frac{i\hbar}{4l^{2}}\int_{0}^{t}\frac{\zeta+1}{m}%
d\tau\right)  , \label{A5}%
\end{equation}
where $C$ is an arbitrary real constant that will be fixed by imposing the
normalization condition on the state $\left\vert \xi,\zeta\right\rangle $.

\section{Normalization condition \label{Appendix-B}}

We will determine the constant $C$ by imposing the normalization of the states
$\left\vert \xi,\zeta\right\rangle $. First, consider the generating function
of the Hermite polynomials, see the formula $\left(  10.13.19\right)  $ in
\citep{ERD1953},
\begin{align}
&  \exp\left(  2yz-z^{2}\right)  =%
%TCIMACRO{\dsum \limits_{n=0}^{\infty}}%
%BeginExpansion
{\displaystyle\sum\limits_{n=0}^{\infty}}
%EndExpansion
\frac{H_{n}\left(  y\right)  }{n!}z^{n},\nonumber\\
&  z=-\sqrt{\frac{\zeta}{2}}\hat{a}^{\dagger},\text{ \ }y=\frac{\xi}%
{\sqrt{2\zeta}}. \label{B1}%
\end{align}
Thus, we can write (\ref{15}), in the form
\begin{equation}
\left\vert \xi,\zeta\right\rangle =C\exp\left(  \frac{1}{2}\frac{\zeta^{\ast
}\xi^{2}}{1-\left\vert \zeta\right\vert ^{2}}-\frac{i\hbar}{4l^{2}}\int
_{0}^{t}\frac{\zeta+1}{m}d\tau\right)
%TCIMACRO{\dsum \limits_{n=0}^{\infty}}%
%BeginExpansion
{\displaystyle\sum\limits_{n=0}^{\infty}}
%EndExpansion
\frac{\left(  -1\right)  ^{n}}{\sqrt{n!}}\left(  \frac{\zeta}{2}\right)
^{\frac{n}{2}}H_{n}\left(  \frac{\xi}{\sqrt{2\zeta}}\right)  \left\vert
n\right\rangle . \label{B2}%
\end{equation}
In what follows, let us calculate $\left\langle \zeta,\xi|\xi,\zeta
\right\rangle $, which will be given by
\begin{equation}
\left\langle \zeta,\xi|\xi,\zeta\right\rangle =\left\vert C\right\vert
^{2}\exp\left[  \frac{\operatorname{Re}\left(  \zeta^{\ast}\xi^{2}\right)
}{1-\left\vert \zeta\right\vert ^{2}}+\frac{\hbar}{2l^{2}}\int_{0}^{t}%
\frac{\operatorname{Im}\left(  \zeta\right)  }{m}d\tau\right]
%TCIMACRO{\dsum \limits_{n=0}^{\infty}}%
%BeginExpansion
{\displaystyle\sum\limits_{n=0}^{\infty}}
%EndExpansion
\frac{\left\vert \zeta\right\vert ^{n}}{2^{n}n!}H_{n}\left(  \frac{\xi}%
{\sqrt{2\zeta}}\right)  H_{n}\left(  \frac{\xi^{\ast}}{\sqrt{2\zeta^{\ast}}%
}\right)  . \label{B3}%
\end{equation}
From the formula (10.13.22) in \citep{ERD1953},
\begin{align}
&
%TCIMACRO{\dsum \limits_{n=0}^{\infty}}%
%BeginExpansion
{\displaystyle\sum\limits_{n=0}^{\infty}}
%EndExpansion
H_{n}\left(  x\right)  H_{n}\left(  y\right)  \frac{z^{n}}{2^{n}n!}=\frac
{1}{\sqrt{1-z^{2}}}\exp\left[  \frac{2xyz-\left(  x^{2}+y^{2}\right)  z^{2}%
}{1-z^{2}}\right]  ,\nonumber\\
&  x=\frac{\xi}{\sqrt{2\zeta}},\text{ \ }y=\frac{\xi^{\ast}}{\sqrt
{2\zeta^{\ast}}},\text{ \ }z=\left\vert \zeta\right\vert , \label{B4}%
\end{align}
and imposing the normalization of the states $\left\vert \xi,\zeta
\right\rangle $, i.e., $\left\langle \zeta,\xi|\xi,\zeta\right\rangle =1$, we
have $C$ in the form
\begin{equation}
C=\left(  1-\left\vert \zeta_{0}\right\vert ^{2}\right)  ^{1/4}\exp\left(
-\frac{1}{2}\frac{\left\vert \xi\right\vert ^{2}}{1-\left\vert \zeta
\right\vert ^{2}}\right)  . \label{B5}%
\end{equation}

\section{Completeness relation \label{Appendix-C}}

Let us get the completeness relation for the states $\left\vert \xi
,\zeta\right\rangle $. As we can see from Eqs. (\ref{11a}) and (\ref{11b}),
the displacement and squeeze parameters are related to each other. Thus, we
will obtain the completeness relation for the SCS by sweeping the entire
complex plane of the constant $\varphi=\varphi_{0}$. Therefore, we must have
\begin{align}
&  \int\left\vert \varphi,\zeta\right\rangle \left\langle \zeta,\varphi
\right\vert d^{2}\varphi=1,\text{ \ }\nonumber\\
&  d^{2}\varphi=\eta\left(  u_{1},u_{2}\right)  du_{1}du_{2},\text{ \ }%
u_{1}=\operatorname{Re}\left(  \varphi\right)  ,\text{ \ }u_{2}%
=\operatorname{Im}\left(  \varphi\right)  . \label{C1}%
\end{align}
where $\eta\left(  u_{1},u_{2}\right)  $ is a weight function, which will be
fixed in order to ensure the completeness relation (\ref{C1}).

Consider the real functions $z_{1}$ and $z_{2}$, in the following form
\begin{equation}
z_{1}=\frac{\cos\left(  \theta_{f}+\frac{1}{2}\theta_{\zeta}\right)
u_{1}+\sin\left(  \theta_{f}+\frac{1}{2}\theta_{\zeta}\right)  u_{2}}%
{\sqrt{2\left\vert \zeta\right\vert }\left\vert f\right\vert },\text{ \ }%
z_{2}=\frac{\cos\left(  \theta_{f}+\frac{1}{2}\theta_{\zeta}\right)
u_{2}-\sin\left(  \theta_{f}+\frac{1}{2}\theta_{\zeta}\right)  u_{1}}%
{\sqrt{2\left\vert \zeta\right\vert }\left\vert f\right\vert }, \label{C2}%
\end{equation}
where we made the substitutions $\zeta=\left\vert \zeta\right\vert
e^{i\theta_{\zeta}}$ and\ $f=\left\vert f\right\vert e^{i\theta_{f}}$. From
here, we can write
\begin{equation}
\left(
\begin{array}
[c]{c}%
z_{1}\\
z_{2}%
\end{array}
\right)  =Z\left(
\begin{array}
[c]{c}%
u_{1}\\
u_{2}%
\end{array}
\right)  ,\text{ \ }Z=\frac{1}{\sqrt{2\left\vert \zeta\right\vert }\left\vert
f\right\vert }\left(
\begin{array}
[c]{cc}%
\cos\left(  \theta_{f}+\frac{1}{2}\theta_{\zeta}\right)  & \sin\left(
\theta_{f}+\frac{1}{2}\theta_{\zeta}\right) \\
-\sin\left(  \theta_{f}+\frac{1}{2}\theta_{\zeta}\right)  & \cos\left(
\theta_{f}+\frac{1}{2}\theta_{\zeta}\right)
\end{array}
\right)  . \label{C3}%
\end{equation}
Thus, the element of integration $d^{2}\varphi$ in the new variables $z_{1}$
and $z_{2}$ takes the form
\begin{equation}
d^{2}\varphi=\eta\left(  z_{1},z_{2}\right)  \det\left(  Z^{-1}\right)
dz_{1}dz_{2}=2\left\vert f\right\vert ^{2}\left\vert \zeta\right\vert
\eta\left(  z_{1},z_{2}\right)  dz_{1}dz_{2}. \label{C4}%
\end{equation}
Therefore, we can rewrite (\ref{C1}), as follows
\begin{align}
&  \int\left\vert \varphi,\zeta\right\rangle \left\langle \zeta,\varphi
\right\vert d^{2}\varphi=2\left\vert f\right\vert ^{2}\left\vert
\zeta\right\vert \sqrt{1-\left\vert \zeta\right\vert ^{2}}%
%TCIMACRO{\dsum \limits_{n,n^{\prime}=0}^{\infty}}%
%BeginExpansion
{\displaystyle\sum\limits_{n,n^{\prime}=0}^{\infty}}
%EndExpansion
\frac{\left(  -1\right)  ^{n+n^{\prime}}\zeta^{\frac{n}{2}}\left(  \zeta
^{\ast}\right)  ^{\frac{n^{\prime}}{2}}}{\sqrt{2^{n+n^{\prime}}n!n^{\prime}!}%
}\left\vert n\right\rangle \left\langle n^{\prime}\right\vert \times
\nonumber\\
&  \int\int\exp\left(  -\frac{2\left\vert \zeta\right\vert }{1+\left\vert
\zeta\right\vert }z_{1}^{2}-\frac{2\left\vert \zeta\right\vert }{1-\left\vert
\zeta\right\vert }z_{2}^{2}\right)  H_{n}\left(  z_{1}+iz_{2}\right)
H_{n^{\prime}}\left(  z_{1}-iz_{2}\right)  \eta\left(  z_{1},z_{2}\right)
dz_{1}dz_{2}. \label{C5}%
\end{align}
Now $\left(  i\right)  $ making the following identification
\begin{equation}
a=\frac{2\left\vert \zeta\right\vert }{1+\left\vert \zeta\right\vert },\text{
\ }b=\frac{2\left\vert \zeta\right\vert }{1-\left\vert \zeta\right\vert },
\label{C6}%
\end{equation}
$\left(  ii\right)  $ considering the orthogonality of the Hermite polynomials
\citep{ISM2015,EIJ1990},
\begin{equation}
\int\int dxdyH_{n}\left(  x+iy\right)  H_{n^{\prime}}\left(  x-iy\right)
e^{-ax^{2}-by^{2}}=\frac{\pi}{\sqrt{ab}}2^{n}n!\left(  \frac{a+b}{ab}\right)
^{n}\delta_{n,n^{\prime}}, \label{C7}%
\end{equation}
and $\left(  iii\right)  $ choosing the weight function $\eta\left(
z_{1},z_{2}\right)  =\left(  \pi\mu\right)  ^{-1}$, we will have
\begin{equation}
\int\left\vert \varphi,\zeta\right\rangle \left\langle \zeta,\varphi
\right\vert d^{2}\varphi=%
%TCIMACRO{\dsum \limits_{n=0}^{\infty}}%
%BeginExpansion
{\displaystyle\sum\limits_{n=0}^{\infty}}
%EndExpansion
\left\vert n\right\rangle \left\langle n\right\vert =1. \label{C8}%
\end{equation}
Thus, we show that the resolution of the identity for the SCS of a free
particle is satisfied in the complex plane of the constant $\varphi$.

\end{document}